\documentclass[conference,authorversion]{IEEEtran}

\usepackage{tikz}
\usepackage{amsmath,amsfonts,bm}
\usepackage{mathtools}
\usepackage{graphicx}
\usepackage{booktabs}
\usepackage{array,graphicx,multirow}
\usepackage{dblfloatfix}
\usepackage{soul}
\usepackage{hyperref}
\usepackage{hhline}
\usepackage{algorithm}
\usepackage{algorithmic}
\usepackage{float}

\let\labelindent\relax 
\usepackage{enumitem}

\usepackage{diagbox}

\usepackage{cancel}

\usepackage{subfig}
\def\removesen{1}
\if 1\removesen
\newcommand\autost[1]{\textcolor{red!40!white}{\sout{#1}}}
\else
\newcommand\autost[1]{}
\fi

\usepackage[normalem]{ulem}
\newcommand\redout{\bgroup\markoverwith
{\textcolor{red}{\rule[.5ex]{2pt}{0.4pt}}}\ULon}
\usepackage{lipsum}
\newcommand\blfootnote[1]{%
  \begingroup
  \renewcommand{\@makefntext}[1]{\noindent\makebox[1.8em][r]#1}
  \renewcommand\thefootnote{}\footnote{#1}%
  \addtocounter{footnote}{-1}%
  \endgroup
}

\usepackage{pifont}

\usepackage{dirtytalk}
\definecolor{cadmiumgreen}{rgb}{0.0, 0.42, 0.24}

\usepackage[framemethod=default]{mdframed}
\newmdenv[linecolor=black,linewidth=1pt,backgroundcolor=gray!20]{myframe}

\usepackage[skins]{tcolorbox}
\newtcolorbox{mytcolorbox}[2][width=0.975\columnwidth,halign=flush center,arc is angular]{colback=gray!10,colframe=black,fonttitle=\bfseries,coltitle=black,colbacktitle=white,enhanced,attach boxed title to top center={yshift=-2mm},title={#2},#1}

\newcommand*\ccfilledred[1]{\tikz[baseline=(char.base)]{
\textcolor{white}{      \node[shape=circle,draw,inner sep=1pt,fill=red!70] (char) {#1};}}}

\definecolor{circlegreencolor}{RGB}{0,124,0}
\newcommand*\ccfilledgreen[1]{\tikz[baseline=(char.base)]{
\textcolor{white}{      \node[shape=circle,draw,inner sep=1pt,fill=circlegreencolor] (char) {#1};}}}

\usepackage{balance}

\usepackage{xcolor}

\newcommand{\myname}{\textit{LLMPirate}}

\newcommand{\AttackGNN}{\textit{AttackGNN}}
\newcommand{\PoisonedGNN}{\texttt{PoisonedGNN}}

\definecolor{mydarkgreen}{RGB}{0, 120, 0}

\usepackage{listings}

\usepackage{url}

\newlength{\oldtabcolsep}
\definecolor{codegreen}{rgb}{0,0.6,0}
\definecolor{codegray}{rgb}{0.5,0.5,0.5}
\definecolor{codepurple}{rgb}{0.58,0,0.82}
\definecolor{backcolour}{rgb}{0.95,0.95,0.92}

\let\othelstnumber=\thelstnumber
\def\createlinenumber#1#2{
    \edef\thelstnumber{%
        \unexpanded{%
            \ifnum#1=\value{lstnumber}\relax
              #2%
            \else}%
        \expandafter\unexpanded\expandafter{\thelstnumber\othelstnumber\fi}%
    }
    \ifx\othelstnumber=\relax\else
      \let\othelstnumber\relax
    \fi
}

\lstdefinestyle{customc}{
  belowcaptionskip=1\baselineskip,
  breaklines=true,
  frame=single,
  xleftmargin=0.35cm,
  xrightmargin=0.15cm,
  numbers=left,
  numbersep=5pt,  
  language=C,
  showstringspaces=false,
  basicstyle=\footnotesize\ttfamily,
  keywordstyle=\bfseries\color{green!40!black},
  commentstyle=\itshape\color{purple!40!black},
  identifierstyle=\color{blue},
  stringstyle=\color{orange},
}

\lstdefinestyle{customcArianeExploit1}{
  breaklines=true,
  frame=single,
  xleftmargin=0.4cm,
  xrightmargin=0.2cm,
  numbers=left,
  numbersep=5pt,  
  language=C,
  showstringspaces=false,
  basicstyle=\footnotesize\ttfamily,
  keywordstyle=\bfseries\color{green!40!black},
  commentstyle=\itshape\color{purple!60!black},
  identifierstyle=\color{blue},
  stringstyle=\color{yellow!50!black},
  morekeywords={asm},
  keywordstyle=[2]\bfseries\color{brown!60!black},
}

\lstdefinestyle{customcArianeExploit}{
  breaklines=true,
  frame=single,
  xleftmargin=0.4cm,
  xrightmargin=0.2cm,
  numbers=left,
  numbersep=5pt,  
  language=C,
  showstringspaces=false,
  basicstyle=\footnotesize\ttfamily,
  keywordstyle=\bfseries\color{blue},
  commentstyle=\itshape\color{green!50!black},
  identifierstyle=\color{black},
  stringstyle=\color{brown},
  morekeywords={asm},
  keywordstyle=[2]\bfseries\color{black},
}

\lstdefinestyle{customlog}{
  breaklines=true,
  frame=single,
  xleftmargin=0.35cm,
  xrightmargin=0.15cm,
  numbers=left,
  numbersep=5pt,  
  language=C,
  showstringspaces=false,
  basicstyle=\footnotesize\ttfamily,
  keywordstyle=\color{blue},
  commentstyle=\itshape\color{purple!40!black},
  identifierstyle=\color{blue},
  stringstyle=\color{orange},
  keywords=[2]{INFO},
  keywords=[3]{ERROR},x
  keywordstyle=[2]\bfseries\color{green!40!black},
  keywordstyle=[3]\bfseries\color{red!500!black},
}

\definecolor{verilogcommentcolor}{RGB}{0,124,0}
\definecolor{verilogkeywordcolor}{RGB}{49,49,255}
\definecolor{backcolor}{RGB}{250,250,250}
\definecolor{verilogsystemcolor}{RGB}{128,0,255}
\definecolor{verilognumbercolor}{RGB}{255,143,102}
\definecolor{verilogstringcolor}{RGB}{160,160,160}
\definecolor{verilogdefinecolor}{RGB}{128,64,0}
\definecolor{verilogoperatorcolor}{RGB}{0,0,128}
\definecolor{pointcolor}{RGB}{192,0,0} 
\lstdefinestyle{prettyverilog}{
   backgroundcolor=\color{backcolor},
   rulecolor=\color{black},
   language           = Verilog,
   commentstyle       = \color{verilogcommentcolor},
   alsoletter         = \$'0123456789\`,
   literate           = *{+}{{\verilogColorOperator{\ +\ }}}{1}%
                         {-}{{\verilogColorOperator{-}}}{1}%
                         {@}{{\verilogColorOperator{@}}}{1}%
                         {;}{{\verilogColorOperator{;}}}{1}%
                         {*}{{\verilogColorOperator{*}}}{1}%
                         {?}{{\verilogColorOperator{? }}}{1}%
                         {:}{{\verilogColorOperator{:}}}{1}%
                         {<}{{\verilogColorOperator{<}}}{1}%
                         {>}{{\verilogColorOperator{>}}}{1}%
                         {!}{{\verilogColorOperator{!}}}{1}%
                         {xorsymbol}{{\verilogColorOperator{^}}}{1}%
                         {|}{{\verilogColorOperator{| }}}{1}%
                         {||}{{\verilogColorOperator{|| }}}{1}%
                         {=}{{\verilogColorOperator{= }}}{1}%
                         {==}{{\verilogColorOperator{== }}}{1}%
                         {=>}{{\verilogColorOperator{=> }}}{1}%
                         {[}{{\verilogColorOperator{[}}}{1}%
                         {]}{{\verilogColorOperator{]}}}{1}%
                         {(}{{\verilogColorOperator{(}}}{1}%
                         {)}{{\verilogColorOperator{)}}}{1}%
                         {rightbracket}{{\verilogColorOperator{)}}}{1}%
                         {,}{{\verilogColorOperator{,}}}{1}%
                         {.}{{\verilogColorOperator{.}}}{1}%
                         {~}{{\verilogColorOperator{$\sim$}}}{1}%
                         {\%}{{\verilogColorOperator{\%}}}{1}%
                         {\&}{{\verilogColorOperator{\& }}}{1}%
                         {\&\&}{{\verilogColorOperator{\&\& }}}{1}%
                         {\#}{{\verilogColorOperator{\#}}}{1}%
                         {\ /\ }{{\verilogColorOperator{\ /\ }}}{3}%
                         {\ _}{\ \_}{2}%
                        ,
   morestring         = [s][\color{verilogstringcolor}]{"}{"},%
   identifierstyle    = \color{black},
   vlogdefinestyle    = \color{verilogdefinecolor},
   vlogconstantstyle  = \color{verilognumbercolor},
   vlogsystemstyle    = \color{verilogsystemcolor},
   basicstyle         = \small\fontencoding{T1}\ttfamily,
  columns=fullflexible, 
   keywordstyle       = \bfseries\color{verilogkeywordcolor},
   morekeywords      = {val, when, port, coverage, unique},
   numbers            = left,
   numbersep          = 5pt,
   tabsize            = 2,
   escapeinside       = {/*!}{!*/},
   upquote            = true,
   sensitive          = true,
   showstringspaces   = false, 
   frame              = single, 
   breaklines         = true,
   abovecaptionskip   = 5pt,
   belowcaptionskip   = 0pt, 
   xleftmargin        =0.35cm,
   xrightmargin       =0.15cm,
   captionpos         = b,
   emph               = {Point, Point0, Point1, Point2, Point3, Point4, Point5, Point6, Point7, Point8, Point9},
   emphstyle          =\color{pointcolor},
   emph               = {[2] STVEC,SCOUNTEREN,MSTATUS,MTVEC,ML1_ICACHE_MISS,ML1_DCACHE_MISS,MITLB_MISS,MDTLB_MISS,
                             MLOAD,MSTORE,MEXCEPTION,MEXCEPTION_RET,MBRANCH_JUMP,MCALL,MRET,MMIS_PREDICT,MSB_FULL,
                             MIF_EMPTY,MHPM_COUNTER_17,MHPM_COUNTER_18,MHPM_COUNTER_19,MHPM_COUNTER_20,MHPM_COUNTER_21,
                             MHPM_COUNTER_22,MHPM_COUNTER_23,MHPM_COUNTER_24,MHPM_COUNTER_25,MHPM_COUNTER_26,MHPM_COUNTER_27,
                             MHPM_COUNTER_28,MHPM_COUNTER_29,MHPM_COUNTER_30,MHPM_COUNTER_31}, 
   emphstyle          = {[2]\bfseries\color{verilogkeywordcolor}}
}

\makeatletter

\newcommand\language@verilog{Verilog}
\expandafter\lst@NormedDef\expandafter\languageNormedDefd@verilog%
  \expandafter{\language@verilog}
  
\lst@SaveOutputDef{`'}\quotesngl@verilog
\lst@SaveOutputDef{``}\backtick@verilog
\lst@SaveOutputDef{`\$}\dollar@verilog

\newcommand\getfirstchar@verilog{}
\newcommand\getfirstchar@@verilog{}
\newcommand\firstchar@verilog{}
\def\getfirstchar@verilog#1{\getfirstchar@@verilog#1\relax}
\def\getfirstchar@@verilog#1#2\relax{\def\firstchar@verilog{#1}}

\newcommand\addedToOutput@verilog{}
\lst@AddToHook{Output}{\addedToOutput@verilog}

\newcommand\constantstyle@verilog{}
\lst@Key{vlogconstantstyle}\relax%
   {\def\constantstyle@verilog{#1}}

\newcommand\definestyle@verilog{}
\lst@Key{vlogdefinestyle}\relax%
   {\def\definestyle@verilog{#1}}

\newcommand\systemstyle@verilog{}
\lst@Key{vlogsystemstyle}\relax%
   {\def\systemstyle@verilog{#1}}

\newcount\currentchar@verilog
  
\newcommand\@ddedToOutput@verilog
{%
   \ifnum\lst@mode=\lst@Pmode%
      \expandafter\getfirstchar@verilog\expandafter{\the\lst@token}%
      \expandafter\ifx\firstchar@verilog\backtick@verilog
         \let\lst@thestyle\definestyle@verilog%
      \else
         \expandafter\ifx\firstchar@verilog\dollar@verilog
            \let\lst@thestyle\systemstyle@verilog%
         \else
            \expandafter\ifx\firstchar@verilog\quotesngl@verilog
               \let\lst@thestyle\constantstyle@verilog%
            \else
               \currentchar@verilog=48
               \loop
                  \expandafter\ifnum%
                  \expandafter`\firstchar@verilog=\currentchar@verilog%
                     \let\lst@thestyle\constantstyle@verilog%
                     \let\iterate\relax%
                  \fi
                  \advance\currentchar@verilog by \@ne%
                  \unless\ifnum\currentchar@verilog>57%
               \repeat%
            \fi
         \fi
      \fi
   \fi
}

\lst@AddToHook{PreInit}{%
  \ifx\lst@language\languageNormedDefd@verilog%
    \let\addedToOutput@verilog\@ddedToOutput@verilog%
  \fi
}

\newcommand{\verilogColorOperator}[1]
{%
  \ifnum\lst@mode=\lst@Pmode\relax%
   {\bfseries\textcolor{verilogoperatorcolor}{#1}}%
  \else
    #1%
  \fi
}

\makeatother

\lstdefinestyle{mystyle}{
    commentstyle=\textit,
    keywordstyle=\textbf,
    stringstyle=\color{codepurple},
    basicstyle=\ttfamily,
    breakatwhitespace=false,         
    breaklines=true,      
    frame=single, 
    abovecaptionskip   = 5pt,
    belowcaptionskip   = 0pt, 
    framexleftmargin=\parindent,
    captionpos=b,                    
    keepspaces=true,                 
    numbers=left,    
    numberstyle=\normalsize,
    stepnumber=1,
    numbersep=5pt,   
    xleftmargin=1.5\parindent,
    showspaces=false,                
    showstringspaces=false,
    showtabs=false,                  
    tabsize=2
}

\lstdefinestyle{normaltext}{
    backgroundcolor=\color{backcolor},
    rulecolor=\color{black},
    commentstyle=\textit,
    basicstyle         = \small\fontencoding{T1}\ttfamily,
    columns=fullflexible, 
    keywordstyle=\textbf,
    numbers            = left,
   numbersep          = 5pt,
   tabsize            = 2,
    showstringspaces   = false,
    breaklines=true,      
    frame=single, 
    abovecaptionskip   = 5pt,
   belowcaptionskip   = 0pt, 
   xleftmargin        =0.35cm,
   xrightmargin       =0.15cm,
   captionpos         = b,
}

\begin{document}

\title{\myname{}: LLMs for Black-box Hardware IP Piracy}
\author{Vasudev Gohil, Matthew DeLorenzo, Veera Vishwa Achuta Sai Venkat Nallam, Joey See, Jeyavijayan Rajendran\\ Electrical and Computer Engineering \\
Texas A\&M University, College Station, Texas \\
{vasudevgohil.personal@gmail.com, \{matthewdelorenzo, nallamsaiv, joeysee,  jv.rajendran\}@tamu.edu} 
}

\maketitle

\begin{abstract}
The rapid advancement of large language models (LLMs) has enabled the ability to effectively analyze and generate code nearly instantaneously, resulting in their widespread adoption in software development. 
Following this advancement,
researchers and companies have also begun integrating LLMs across the hardware design and verification process.
However, these highly potent LLMs can also induce new attack scenarios upon security vulnerabilities across the hardware development process. One such attack vector that has not been explored so far is intellectual property (IP) piracy. Given that this attack can manifest as rewriting hardware designs to evade 
piracy detection, it is essential to thoroughly evaluate LLM capabilities in performing this task and assess the mitigation abilities of current IP piracy detection tools.

Therefore, in this work, we propose \myname{}, the first LLM-based technique able to generate pirated variations of circuit designs that successfully evade detection across multiple state-of-the-art piracy detection tools. 
We devise three solutions to overcome challenges related to integration of LLMs for hardware circuit designs, scalability to large circuits, and effectiveness, resulting in an end-to-end automated, efficient, and practical formulation.
We perform an extensive experimental evaluation of \myname{} using eight LLMs of varying sizes and capabilities and assess their performance in pirating various circuit designs against four state-of-the-art, widely-used piracy detection tools. 
Our experiments demonstrate that \myname{} is able to consistently evade detection on 100\% of tested circuits across every detection tool. Additionally, we showcase the ramifications of \myname{} using case studies on \texttt{IBEX} and \texttt{MOR1KX} processors and a \texttt{GPS} module, that we successfully pirate.
We envision that our work motivates and fosters the development of better IP piracy detection tools.
\end{abstract}

\maketitle

\section{Introduction}\label{sec:introduction}
Recent advancements within artificial intelligence and computing performance have greatly accelerated the development of large language models (LLMs), with state-of-the-art models (including OpenAI's ChatGPT~\cite{openai_intro} and Google's Bard~\cite{Pichai_2023}) achieving groundbreaking performance in natural language processing and gaining mass popularity~\cite{reuters_Hu}.
With the ability to effectively interpret text prompts and generate human-like responses~\cite{naveed2023comprehensive}, LLMs have proven effective across a variety of tasks, such as language translation~\cite{kocmi2023large}, text summarization~\cite{pandya2023automating}, and generating code~\cite{fan2023large}. This widespread applicability has resulted in the rapid adoption of LLMs across various industries, serving as chat-bots for customer service~\cite{pandya2023automating}, documentation aids in healthcare~\cite{Meskó_Topol_2023}, and coding assistants for programmers~\cite{fan2023large}.
These applications have prompted companies and researchers to further explore the most effective ways in which LLMs can be tailored and utilized to automate specified tasks and processes, including the software and hardware design workflow.

\subsection{LLMs for Code Generation}
Given the success of LLMs in natural language processing, many models are also extensively trained on large datasets of open-source code with the specific purpose of generating functionally correct programs based upon a prompt description~\cite{codellama_announcement}. These programming-oriented LLMs are utilized in a variety of applications within the software and hardware development processes. Through Microsoft's Github CoPilot, the advantages of LLMs are applied directly to the software development environment, providing context-aware code suggestions and refactoring recommendations~\cite{GitHub_features}. In fact, Microsoft reported that the first versions of CoPilot tools substantially
increase productivity on common enterprise information worker tasks~\cite{cambon2023early}. Similarly, OpenAI's widely utilized GPT-4 model also has strong performance in software engineering tasks, including the ability to generate programs from pseudocode and explain its results in natural language~\cite{bubeck2023sparks}.

Following advancements in the software domain, semiconductor companies have also begun utilizing generative artificial intelligence (AI), specifically LLMs, within various stages of hardware integrated circuit design process, including the generation of register-transfer level (RTL) code. 
Semiconductor giant NVIDIA's ChipNeMo explores fine-tuning smaller LLMs for industrial chip-design, in which their 70-billion parameter model was able to outperform OpenAI's GPT-4 in electronic design automation (EDA) tools' script generation~\cite{liu2024chipnemo}.
ChipGPT from Cadence demonstrated the first proof-of-concept LLM technology in chip design, able to load architecture and design specifications to accelerate test-bench creation and RTL code generation~\cite{Brown}. Cadence has also developed the Cadence.AI generative AI platform for applications in digital circuit design, analog circuit design, debug and verification, and printed circuit board design~\cite{cadence_ai}. Likewise, Synopsys, another EDA enterprise, has developed Synopsys.ai Copilot that harnesses generative AI with LLMs throughout their EDA suite, aiding in tedious workflow tasks including test pattern generation, verification coverage, and design space exploration~\cite{Rangarajan_2023}. RapidGPT from Rapid Silicon provides similar auto-complete capabilities tailored to field-programmable gate array design~\cite{CorporateTeam_2023}.

LLMs can also be offensively leveraged by threat actors to execute attacks that exploit various software and hardware security vulnerabilities. For instance, RatGPT utilizes GPT-4 as a proxy method to distribute malicious software through the use of openly accessible LLM plugins, enabling access to the victim's machine~\cite{beckerich2023ratgpt}. Additionally, GPThreats-3~\cite{10188649} explores how LLMs (GPT-3) can be utilized to generate malware itself, demonstrating success through a building-block prompting strategy. To attack hardware systems, LLMs can enhance side-channel attacks, in which unintentional data is extracted based upon the physical implementation of the circuit design (e.g., the secret key from a cryptographic algorithm). AgentSCA demonstrates that through fine-tuning LLMs with human feedback upon side-channel datasets, the model can effectively interpret side-channel statistics and provide correct decisions based upon the test system~\cite{llm_sidechannel}. 
However, \textbf{a crucial attack vector that can be orchestrated using LLMs that has not been explored so far is the piracy of 
intellectual property (IP) within the hardware domain.}
\begin{figure}
    \centering
    \includegraphics[width=0.48\textwidth,trim={0.0cm 0.2cm 0.2cm 0.0cm},clip]{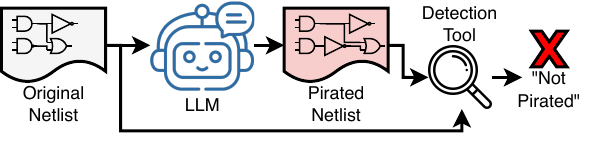}
    \caption{High-level overview of our proposed technique.}
    \label{fig:figure1_overview}
\end{figure}

\subsection{Impact of IP Piracy}
The theft of hardware design IP, or IP piracy, is a significant concern within the system-on-chip design flow~\cite{5401214}. 
This can be attributed to the globalization of the integrated circuits supply chain, where semiconductor companies outsource their IP design to (potentially untrusted) fabrication entities to reduce the cost and time of chip production~\cite{hassija2020survey}. This has caused an increased risk of theft of IP assets shared by vendors (including RTL designs), causing significant security and economic consequences. A recent instance is observed within the dynamic random-access memory (DRAM) market. Micron, who held 20-25\% of the global DRAM market share, reported estimated losses of \$8.75 billion to IP piracy alone in 2018, demonstrating significant economic impacts~\cite{sessions2018attorney}. Additionally, the semiconductor industry has been largely
impacted, with the U.S. Trade Representative reporting losses between \$225 to \$600 billion as a result of Chinese theft of American IP~\cite{UnitedStates, Cohen_2019, Microsoft_ip_piracy}. 

To address
this threat across the hardware design process, a number of hardware IP protection techniques~\cite{protect_IP_overview} as well as piracy detection tools such as GNN4IP are utilized~\cite{GNN4IP}.
However, piracy detection tools such as GNN4IP have not been thoroughly tested. In this work, we show how LLMs can be used to pirate IPs and successfully evade tools such as GNN4IP.

\begin{table*}[t]
\centering
\caption{Overview of \myname{} against piracy detection tools.}
\resizebox{\textwidth}{!}{
\begin{tabular}{ccccc}
\toprule
\textbf{Detection Tool} & GNN4IP~\cite{GNN4IP} & MOSS~\cite{moss_website} & Jplag~\cite{Jplag_github} & SIM~\cite{SIM_website} \\
\cmidrule{1-5}
\textbf{Algorithm Used} & \begin{tabular}[c]{@{}c@{}} Graph Neural Network\end{tabular}  & Winnowing & Greedy String Tiling & \begin{tabular}[c]{@{}c@{}}Tokenization and\\Longest Common Subsequence\end{tabular} \\
\cmidrule{1-5}
\textbf{Key Features} & \begin{tabular}[c]{@{}c@{}}High Accuracy,\\Designed for Verilog\end{tabular} & \begin{tabular}[c]{@{}c@{}}Widely-used,\\Supports Verilog\end{tabular} & \begin{tabular}[c]{@{}c@{}}Widely-used,\\Robust Against Obfuscation\end{tabular} & \begin{tabular}[c]{@{}c@{}}Widely-used,\\Efficient\end{tabular}\\ \cmidrule{1-5}
\textbf{\begin{tabular}[c]{@{}c@{}}\myname~(This Work)'s\\ Evasion Rate\end{tabular}} & $100$\% & $100$\% & $100$\% & $81.25$\%\footnotemark  \\
\bottomrule
\end{tabular}
}
\label{tab:overview_table}
\end{table*}

\subsection{Our Goals and Contributions}
We propose an end-to-end automated LLM-based IP piracy scheme, \myname{}, using which we can rewrite circuits, i.e., Verilog netlists, such that they evade detection by various IP piracy detection tools. Figure~\ref{fig:figure1_overview} illustrates the high-level idea.
This requires designing an appropriate task for a given LLM, such that prompting it with a target netlist results in a design that is functionally equivalent to the original circuit, but is also different enough to not be flagged by piracy detection tools.

However, several challenges exist in designing such an end-to-end automated IP piracy framework. First, LLMs are trained on extremely limited hardware designs, i.e., circuits described in hardware description languages such as Verilog or VHDL, as opposed to software codes such as C, C++, Java, or Python~\cite{thakur2023verigen}. In fact, our experiments indicate that even advanced LLMs such as OpenAI's GPT-3.5 are unable to understand and rewrite simple Verilog designs. Second, these LLMs face scalability issues: they are unable to work with larger Verilog designs. Another scalability issue also stems from the limited context windows (i.e., the amount of information that an LLM can take into account to generate responses without losing context) of LLMs. This is especially important since typical Verilog designs have hundreds of thousands of characters, which is far more than the context windows of all LLMs available today. Third, LLMs' responses are not deterministic, so often, simply prompting an LLM to rewrite/pirate circuit designs, i.e., netlists, results in error-prone/incorrect responses. This needs to be alleviated in order to realize a practical IP piracy technique.

We overcome the challenge of limited Verilog data used to train LLMs by devising Solution \ccfilledgreen{A}: translating the syntax of Verilog netlists to a more generic format of Boolean functions before prompting the LLMs. Doing so makes it easier for LLMs to understand the prompt details and respond accordingly. To overcome the challenges related to scalability, we devise Solution \ccfilledgreen{B}, which characterizes large netlists to extract all unique gate types and then using a divide-and-conquer approach to not exceed the context window sizes of LLMs. Finally, we overcome the challenge related to error-prone responses, we devise Solution \ccfilledgreen{C}, which combines the interactive capabilities of LLMs with fine-grained feedback related to syntactical and functional correctness of the circuit generated by the LLM. Sec.~\ref{sec:methodology_pirating_firm_ips} contains more details about these challenges and our solutions. By solving these challenges, we develop an end-to-end automated technique, \myname{}, for successfully pirating hardware IPs. Our primary contributions are:

\begin{itemize}[leftmargin=*]
\item We present a first-of-its-kind LLM-based technique, \myname{}, that successfully pirates hardware IPs and evades detection by state-of-the-art detection tools. This is also the first work to provide a detailed comparative study of the efficacy of four different piracy detection tools that use different algorithms.
\item We overcome unique challenges related to integration of LLMs for Verilog netlists, scalability, and effectiveness by designing custom solutions based on domain knowledge.
\item We provide a comparative evaluation between various popular LLM models across various IP piracy detection tools, with results indicating that overall, GPT-4 is the most effective in successfully pirating Verilog netlists. Other large closed-source LLMs such as CoPilot and GPT-3.5 also perform very well.
\item Our results also demonstrate that our feedback-guided interactive formulation greatly improves the performance of all LLMs, and most notably of smaller, open-source LLMs such as the recently released Llama3.
\item We demonstrate the practical ramifications of our LLM-based IP piracy technique through case studies on modern, real-world designs: \texttt{IBEX} and \texttt{MOR1KX} processors and a \texttt{GPS} module. We successfully pirate them and evade multiple state-of-the-art detection tools.
\end{itemize}

\subsection{Why LLMs?}
A natural question here could be about the need of LLMs for hardware IP piracy. LLMs have shown tremendous improvement over the past few years, and today's LLMs are proficient at a variety of tasks including programming~\cite{jiang2024survey}. However, as we demonstrate in Sec.~\ref{sec:methodology_pirating_firm_ips}, LLMs still struggle with understanding simple Verilog netlists.
Given this limitation of LLMs, a potential approach for a malicious human developer can be to manually rewrite netlists.
However, such an approach would not be scalable. Additionally, the limitations of this manual approach are exacerbated by the fact that different piracy detection techniques use different types of algorithms (e.g., graph structural similarity,  ``fingerprints'' of hashed Verilog netlist structures, and text-based comparisons). Another limitation of the manual approach is the requirement of additional effort for every new piracy detection technique. 
In contrast, \myname{} offers an end-to-end automated flow to easily and quickly pirate hardware circuit netlists, enabling proper evaluation of existing and new piracy detection techniques.

\footnotetext{\myname{} does not have 100\% evasion rate against SIM because SIM has high false-positive rate as it does not support Verilog natively.}
\section{Background}\label{sec:background}

\subsection{Large Language Models}
In recent years, large language models (LLMs) have emerged as powerful tools in natural language processing and related fields. These models, often based on deep learning architectures, exhibit remarkable capabilities in tasks such as text generation, translation~\cite{kocmi2023large}, and sentiment analysis~\cite{krugmann2024sentiment}. LLMs learn to represent language patterns and context by training on massive amounts of text data. Notable examples include GPT-3.5 (used in ChatGPT)~\cite{chatgpt_website}, GPT-4~\cite{gpt4_website}, Gemini~\cite{gemini_website}, and Llama~\cite{llama3_website}, among others. The remarkable success of LLMs can be attributed to their architectural innovation, in which 
transformer architectures are leveraged to enable
parallel processing of sequential data, thereby efficiently capturing complex linguistic patterns and dependencies within text~\cite{vaswani2017attention}. 

\subsection{Code Generation with Large Language Models}
Among other avenues, LLMs have also brought a paradigm shift in code generation~\cite{jiang2024survey}. LLMs have demonstrated remarkable proficiency in generating code across different programming languages~\cite{cassano2023multipl}. By leveraging the vast knowledge encoded in their pre-trained parameters, these models can understand natural language prompts describing desired functionalities or requirements and produce corresponding code snippets with high fidelity. This capability has shown promise for accelerating software development processes, facilitating rapid prototyping, and reducing the burden on programmers by automating routine coding tasks~\cite{AI_routine_coding,GitHub_features}. Additionally, fine-tuning these models on domain-specific codebases further enhances their proficiency in generating contextually relevant and syntactically correct code~\cite{roziere2023code}. For instance, researchers and corporations have devised custom LLMs for generating codes in hardware description languages, such as Verilog and VHDL, which are used to create digital integrated circuits~\cite{thakur2023verigen,fu2024hardware}.

\subsection{IP Piracy Detection}\label{sec:background_piracy_detection}

Although there are many noteworthy works for measuring similarity, we choose four techniques for our evaluation, as explained next. 
Our selection of target similarity detection techniques ranges from the earliest tools
with high popularity, MOSS~\cite{moss_website}, to the most recent, GNN4IP~\cite{GNN4IP}, which uses machine learning. We also select other tools, SIM~\cite{SIM_website} and Jplag~\cite{Jplag_github}, based on their high accuracy, impact, and popularity (see Table~\ref{tab:overview_table} for an overview).
MOSS is arguably the most widely-used similarity measurement tool for codes. It has been used globally for decades~\cite{sheahen2016taps,simon2020choosing,bowyer1999experience,Monash_using_MOSS_and_Jplag}, has over 300K active accounts~\cite{mossad}, and is also used in (or for the basis of) commercial tools for similarity detection, such as Gradescope~\cite{Gradescope_using_MOSS} and Codequiry~\cite{codequiry_using_MOSS}.
GNN4IP is the most powerful similarity measurement tool for Verilog, as it was developed with the specific objective of detecting IP piracy in Verilog code.
Jplag, like MOSS, is also a widely-used similarity measurement tool that is used in universities~\cite{Monash_using_MOSS_and_Jplag}.
Overall, our selection represents a set of similarity detection tools that use different frameworks, demonstrate
excellent performance,
and have been used extensively.

\noindent\textbf{GNN4IP} is a piracy detection tool developed
with the specific purpose of detecting hardware IP piracy~\cite{GNN4IP}. It converts Verilog descriptions of hardware IP into graph representations and performs graph convolutions on those graphs in order to extract their node embeddings. By finding the cosine similarity between the embeddings of different IPs, it then becomes possible to determine if IP piracy has occurred.

\noindent\textbf{MOSS}
is a piracy detection tool developed by Stanford University~\cite{moss_website}, primarily utilized for detecting plagiarism in code across students in college-level computer science courses. MOSS uses the winnowing algorithm~\cite{schleimer2003winnowing}, which first breaks down code into tokens and hashes them using a hash function, then applies a sliding window over the hashes, and lastly selects the minimum hash value from each window. These values become the ``fingerprint'' of the code, which are then utilized to evaluate the similarity percentage between two target codes.

\noindent\textbf{JPlag}
is a Java-based software similarity detection tool~\cite{Jplag_github}.
It utilizes the Greedy-String-Tiling algorithm on tokenized entries to systemically break entries up into ``tiles'' of matching token strings, prioritizing the longest strings first. The number of tiles found are then compared to the overall length of entries to assess similarities. Originally developed in 1996, it has JPlag has received consistent updates and improvements by universities and other community members who continue to use it today~\cite{Monash_using_MOSS_and_Jplag}.

\noindent\textbf{SIM} is another piracy detection tool, utilized to detect plagiarism in writing assignments and software projects across college students~\cite{SIM_website}. SIM firstly breaks the given code/text into 16-bit tokens using a hash function and normalizes them in order to minimize superficial differences (such as comments, white spaces, etc). Then the algorithm scans for overlapping blocks of tokens that appear in both files.
Finally, the similarity percentage is found by dividing the number of matching tokens by the total number of tokens~\cite{SIM_manual}.
\section{Threat Model}\label{sec:threat_model}

We consider a standard black-box attacker model applicable to piracy detection or similarity measurement techniques such as GNN4IP, MOSS, Jplag, and SIM. In this context, we establish the following assumptions about the attacker:

\noindent\textbf{Attacker’s Knowledge.} We assume a black-box setting, where the attacker lacks access to the detection tool's internal parameters (e.g., ML model's parameters or training labels, or internal parameters used in algorithms of the detection tool). The attacker can only make black-box queries to obtain output similarity scores or predicted labels (in case of ML-based techniques).

\noindent\textbf{Attacker’s Capacity.} The piracy attack occurs after the detection tool is finalized. Especially in case of machine learning (ML)-based techniques, the attack occurs after the model has undergone training. The detection tool remains fixed, and the adversary lacks the ability to alter its parameters or structure. For instance, the attacker cannot introduce model poisoning (for ML-based techniques) or inject backdoors.

\noindent\textbf{Attacker’s Abilities.} The attacker can rewrite the netlist arbitrarily, but not alter the netlist's functionality. Additionally, the attacker must adhere to circuit design rules.

\noindent\textbf{Attacker’s Goal.} The attacker's objective is to generate netlists that lead to misclassification by the target detection tool(s). For instance, when the target detection tool is GNN4IP, the attacker aims to create a pirated version of an original netlists such that GNN4IP incorrectly classifies the pirated netlist as “not pirated”. Or, when the target detection tool is MOSS, the attacker aims to pirate an original netlist such that MOSS returns a low enough similarity score (determined by a threshold, explained in Sec.~\ref{sec:results}).

\section{Methodology}\label{sec:methodology}
In this section, we first provide a preliminary formulation to pirate firm hardware intellectual property (IP), i.e., gate-level Verilog netlists, using LLMs. Then, we show that this preliminary formulation only works to an extent and doesn't help us successfully pirate IPs. Then we delve into the details of the limitations and describe the different challenges that need to be overcome to achieve our goal. We also devise solutions to address these challenges and build our framework, which successfully pirates hardware IP and evades detection by state-of-the-art piracy detection tools solely through black-box LLM access.

\subsection{LLMs for Pirating IPs - Formulation, Challenges, and Solutions}\label{sec:methodology_pirating_firm_ips}
Here, we devise a preliminary formulation using LLMs to pirate IPs in the form of gate-level netlists.
To that end, consider the example prompt shown in Listing~\ref{listing:prelim_formulation_prompt}. Here, we simply ask the LLM to rewrite a Verilog gate-level netlist for an \texttt{OR} gate.
Mathematically, this formulation can be represented as follows: The response $R$ is obtained from the underlying distribution 
\[
p(R|Q,\theta), \tag{1}\label{eq:prelim_formulation}
\]
where $\theta$ denotes the parameters of the LLM, $Q$ denotes the query, and $p$ represents the probability (since LLMs' responses are not deterministic).

\begin{lstlisting}[language=Verilog, label = {listing:prelim_formulation_prompt}, caption={Prompt for rewriting Verilog netlist.}, style=prettyverilog, float, belowskip=-15pt, aboveskip=10pt, firstnumber=1, linewidth=\linewidth]
// Can you refactor it so that functionality remains the same, but the gates used and their interconnections are different from the original structure?
module top (input a, input b, output crightbracket;
or U1 (c, a, brightbracket;
endmodule
\end{lstlisting}

\begin{lstlisting}[language=Verilog, label = {listing:prelim_formulation_response}, caption={GPT-3.5's response to the prompt in Listing~\ref{listing:prelim_formulation_prompt}.}, style=prettyverilog, float, belowskip=-15pt, aboveskip=20pt, firstnumber=1, linewidth=\linewidth]
module top (input a, input b, output crightbracket;
nand n1 (w1, a, brightbracket;
nand n2 (c, w1, w1rightbracket;
endmodule
\end{lstlisting}

Listing~\ref{listing:prelim_formulation_response} contains the code portion of an example response from the LLM.\footnote{Although the listing shows GPT-3.5's response, we also tested GPT-4 and observed similar results.} The generated netlist is a valid gate-level netlist, but it is not functionally equivalent to the original
(as it implements an \texttt{AND} gate, not an \texttt{OR} gate). Similar results hold true for other simple netlists as well, which leads us to the first challenge in pirating IPs.

\noindent\textbf{Challenge \ccfilledred{1}: Difficulty Understanding and Rewriting Simple Hardware Circuit Netlists.}
Although LLMs understand the syntax of gate-level netlists and generate syntactically correct netlists that compile successfully, the generated netlists do not maintain the same functionality even for extremely small and simple modules. This is likely
because most of the Verilog codes available on GitHub and other sources for LLMs' training data is at the RTL abstraction and not at the gate-level netlist abstraction.

\noindent\textbf{Solution \ccfilledgreen{A}: Prompt Syntax Translation For Hardware Netlists.}\label{sec:solution_1} To address this challenge, we revise the formulation to (i)~extract only the relevant parts (i.e., the gates and not the \texttt{module} declarations, \texttt{endmodule} declaration, etc.) from the Verilog netlist, and (ii)~translate the syntax of the extracted gates into a more generic format of Boolean functions (as opposed to gate declarations in the standard Verilog syntax). For instance, the standard Verilog syntax of ``\texttt{or U1 (c, a, b);}'' would be translated into a generic Boolean function format as ``\texttt{c = OR(a,b)}''. Mathematically, in this updated formulation, the response $R$ is obtained from the underlying distribution 
\[
p(R| \mathcal{T}(Q),\theta), \tag{2}\label{eq:solution1}
\]
where $\mathcal{T}(Q)$ denotes that the query, $Q$, is processed to extract the relevant parts (i.e., the gates). These gates are then translated (denoted by $\mathcal{T}(\cdot)$) into a generic Boolean function format, which assists the LLM in generating better responses. Finally, note that the response $R$ is also post-processed using $\mathcal{T}^{-1}$, the inverse of $\mathcal{T}$, to translate the generic Boolean function syntax back to the standard Verilog syntax. We omit this in the formulation for the sake of clarity. For additional information regarding the translation process, see Sec.~\ref{sec:appendix_conversion_from_boolean_formulas} of the Appendix.

Listings~\ref{listing:solution1_prompt} and~\ref{listing:solution1_response} show the translated prompt, $\mathcal{T}(Q)$, according to this updated formulation and the corresponding response, $R$, from GPT-3.5, respectively. As shown, the LLM is not only able to understand the provided circuit in the generic Boolean function format, but it actually rewrites the circuit correctly using the \texttt{NAND} and \texttt{NOT} Boolean functions while maintaining the overall functionality. Thus, theoretically, LLMs can be used to modify gates with the objective of evading evade piracy detection tools. However, in practice, when we use the above formulation (i.e., the one in Eq.~(\ref{eq:solution1})), we face challenges related to scalability and effectiveness. Next, we describe these challenges and how we overcome them.

\begin{lstlisting}[language={}, label = {listing:solution1_prompt}, caption={Updated prompt for rewriting Verilog netlist.}, style=normaltext, float, belowskip=-15pt, aboveskip=10pt, firstnumber=1, linewidth=\linewidth]
Can you refactor this circuit so that functionality remains the same but the Boolean operators are different? Return the circuit in the same format.
c = OR(a,b)
\end{lstlisting}

\begin{lstlisting}[language={}, label = {listing:solution1_response}, caption={GPT-3.5's response to the prompt in Listing~\ref{listing:solution1_prompt}.}, style=normaltext, float, belowskip=-15pt, aboveskip=20pt, firstnumber=1, linewidth=\linewidth]
c = NAND(NOT(a), NOT(b))
\end{lstlisting}

\noindent\textbf{Challenge \ccfilledred{2}: Lack of Scalability to Large Netlists.} Although the LLM successfully rewrites the netlist in the example above, that example contains a toy netlist with just one gate. Real-world netlists contain several thousands, if not hundreds of thousands, of gates. To check the formulation's capability in scaling to larger netlists, we test for a small standard benchmark circuit, \texttt{c17}, which contains 6 gates, as shown in Listing~\ref{listing:challenge2_circuit}. Following the formulation in Eq.~(\ref{eq:solution1}), we query the LLM with the prompt shown in Listing~\ref{listing:challenge2_prompt}. The LLM's response in Listing~\ref{listing:challenge2_response} shows that 
although it follows the instruction and uses different Boolean operators, the resulting circuit is not functionally equivalent to the original circuit (because \texttt{AND(NOT(a), NOT(b))} $\neq$ \texttt{NAND(a, b)}).

\begin{lstlisting}[language=Verilog, label = {listing:challenge2_circuit}, caption={\texttt{c17} benchmark Verilog netlist.}, style=prettyverilog, float, belowskip=-15pt, aboveskip=10pt, firstnumber=1, linewidth=\linewidth]
module c17 (N1,N2,N3,N6,N7,N22,N23rightbracket;
input N1,N2,N3,N6,N7;
output N22,N23;
wire N10,N11,N16,N19;
nand U1 (N10, N1, N3rightbracket;
nand U2 (N11, N3, N6rightbracket;
nand U3 (N16, N2, N11rightbracket;
nand U4 (N19, N11, N7rightbracket;
nand U5 (N22, N10, N16rightbracket;
nand U6 (N23, N16, N19rightbracket;
endmodule
\end{lstlisting}

\begin{lstlisting}[language={}, label = {listing:challenge2_prompt}, caption={Prompt corresponding to Listing~\ref{listing:challenge2_circuit}.}, style=normaltext, float, belowskip=-15pt, aboveskip=10pt, firstnumber=1, linewidth=\linewidth]
Can you refactor this circuit so that functionality remains the same but the Boolean operators are different? Return the circuit in the same format.
N10 = NAND(N1, N3)
N11 = NAND(N3, N6)
N16 = NAND(N2, N11)
N19 = NAND(N11, N7)
N22 = NAND(N10, N16)
N23 = NAND(N16, N19)
\end{lstlisting}

\begin{lstlisting}[language={}, label = {listing:challenge2_response}, caption={GPT-3.5's response to the prompt in Listing~\ref{listing:challenge2_prompt}.}, style=normaltext, float, belowskip=-15pt, aboveskip=0pt, firstnumber=1, linewidth=\linewidth]
N10 = AND(NOT(N1), NOT(N3))
N11 = AND(NOT(N3), NOT(N6))
N16 = AND(NOT(N2), NOT(N11))
N19 = AND(NOT(N11), NOT(N7))
N22 = AND(NOT(N10), NOT(N16))
N23 = AND(NOT(N16), NOT(N19))
\end{lstlisting}

\noindent\textbf{Challenge \ccfilledred{3}: Limited Token Context Windows of LLMs.} Another limitation of the above formulation is that rewriting netlists by simply providing all gates to the LLM is not possible. This is because all LLMs have finite input token context windows, meaning that the prompt size cannot be too large. For example, OpenAI's GPT-3.5 LLM (more specifically, \texttt{gpt-3.5-turbo-0125}) has a context window of 16,385 tokens~\cite{gpt3dot5_context_window}, which, assuming $\approx$4 characters per token~\cite{tokens_to_chars}, translates to $\approx$65,540 characters.
However, practical netlists containing thousands or more gates have hundreds of thousands of characters. Thus, it is not possible to rewrite realistic netlists by providing all gates to LLMs.

\begin{table*}[t]
\caption{Allowed Boolean operators for different gates crafted to achieve structural differences.}
\label{tab:allowed_boolean_operators}
\resizebox{\textwidth}{!}{
\begin{tabular}{ccccccc}
\toprule
\textbf{Gate} & \texttt{AND} & \texttt{OR} & \texttt{NAND} & \texttt{NOR} & \texttt{XOR} & \texttt{XNOR} \\ \midrule
\begin{tabular}[c]{@{}c@{}} \textbf{Allowed Operators}\\\textbf{For Transformation}\end{tabular} & 
[\texttt{NAND}]/[\texttt{NOR}]/[\texttt{OR}, \texttt{NOT}] & 
[\texttt{NAND}]/[\texttt{NOR}]/[\texttt{AND}, \texttt{NOT}] & 
[\texttt{NOR}]/[\texttt{AND}, \texttt{NOT}]/[\texttt{OR}, \texttt{NOT}] & 
[\texttt{NAND}]/[\texttt{AND}, \texttt{NOT}]/[\texttt{OR}, \texttt{NOT}] & 
[\texttt{NAND}]/[\texttt{NOR}] & 
[\texttt{NAND}]/[\texttt{NOR}]
\\ \bottomrule
\end{tabular}
}
\end{table*}

\noindent\textbf{Solution \ccfilledgreen{B}: Pre-characterization and Divide-and-conquer.}
To address these challenges, we modify the formulation by characterizing the given netlist, as explained next. Suppose we wish to rewrite a given Verilog gate-level netlist. Instead of simply extracting all the gates and translating them to create one big prompt (as shown in Listings~\ref{listing:challenge2_circuit} and~\ref{listing:challenge2_prompt}), we first analyze the netlist and extract all the different gate types (e.g., 2-input \texttt{AND} gates, 3-input \texttt{AND} gates, \texttt{XOR} gates, etc.). Then, for each unique gate type, we create a representative circuit in a generic Boolean function format, as explained in Solution \ccfilledgreen{A} above. Finally, for each representative circuit in the generic Boolean function format, we independently prompt the LLM to rewrite that circuit. Note that, for each representative circuit (i.e., gate type), we devise lists of specific Boolean operators (different from the gate in the original circuit) 
and instruct the LLM to only use operators
from that list 
in order to achieve structural differences and result in successful piracy. For instance, if the representative circuit is for an \texttt{OR} gate, the list of Boolean operators
the LLM is allowed to use is one of the following: [\texttt{NAND}], [\texttt{NOR}], or [\texttt{AND}, \texttt{NOT}].\footnote{We create these lists to ensure that operators in each list form a ``complete set'', i.e., it is possible to rewrite the given representative circuit gate using only the operators in the list.} Table~\ref{tab:allowed_boolean_operators} shows the different Boolean operators we allow for each type of gate. Also, Listing~\ref{listing:solution2_prompt} shows an example prompt when we want the LLM to rewrite a 2-input \texttt{AND} gate. From the three available operator options for the \texttt{AND} gate, in this example, we choose the [\texttt{OR}, \texttt{NOT}] operators. The last two instructions in the prompt are provided for ease of parsing the generated response.

\begin{lstlisting}[language={}, label = {listing:solution2_prompt}, caption={Example prompt for instructing the LLM to use only specific Boolean operators from Table~\ref{tab:allowed_boolean_operators} according to Solution {\protect\ccfilledgreen{B}.}}, style=normaltext, float, belowskip=-15pt, aboveskip=10pt, firstnumber=1, linewidth=\linewidth]
Can you refactor this circuit following these instructions? 1) Use only OR and/or NOT  Boolean operators. 2) Ensure that the final functionality remains the same. 3) Just give me the new circuit and nothing else. 4) Generate your response in the following format: <output> = <gate type> (<inputs>).
c = AND(a,b)
\end{lstlisting}

In essence, we use a divide-and-conquer approach where instead of asking the LLM to rewrite the entire netlist, we analyze the netlist, extract different types of gates, and then ask the LLM to individually rewrite the circuits corresponding to the different types of gates. Doing so (i)~allows us to successfully scale to large netlists since we characterize them and focus on different gate types individually, and (ii)~overcomes the issue of limited token context windows of LLMs since the prompt for each unique gate type is independent of others. Mathematically, the formulation is updated to obtain response $R_i$ for the $i^{\text{th}}$ unique gate type from the underlying distribution
\[
p(R_i| \mathcal{T}(Q_i),\theta), \quad \forall i \in \{1,2,\ldots,|G|\}, \tag{3}\label{eq:solution2}
\]
where $G$ is the set of unique gate types and $\mathcal{T}(Q_i)$ denotes the translated query (from Verilog gate format to generic Boolean function format with instructions about allowed Boolean operators) for the $i^{\text{th}}$ unique gate type.

\noindent\textbf{Challenge \ccfilledred{4}: Error-prone Single-shot Netlists.} The final issue with the formulation described so far is that it only allows the LLMs one chance to generate a functionally equivalent circuit that uses different Boolean operators. However, due to randomness in the LLMs' responses and differing training processes (including number of parameters, size and quality of training data, and training processes such as pre-training, fine tuning, instruction tuning, reinforcement learning, etc.), different LLMs have varying amounts of success in rewriting circuits. This implies that, with this single-shot formulation, an LLM's response could be classified as a failure even though it might only generate a slightly incorrect circuit.

\begin{figure*}[t]
    \centering
    \includegraphics[width=\textwidth,trim={0.0cm 0.0cm 0.0cm 0.0cm},clip]{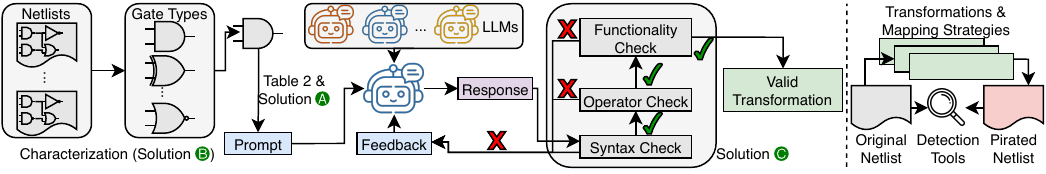}
    \caption{\myname{}'s end-to-end automated flow. All steps, including characterization, prompt syntax translation, syntax, operator and functionality checks, feedback, and the generation of pirated netlists using the LLM-generated transformations are end-to-end automated, and no manual intervention is needed.}
    \label{fig:final_flow}
\end{figure*}

\noindent\textbf{Solution \ccfilledgreen{C}: Feedback-guided Interactive Formulation.} To overcome this issue and ensure that LLMs are not penalized for minor mistakes, we leverage the interactive capabilities of LLMs by combining them with multi-level fine-grained feedback. More specifically, we allow the target LLM $M$ attempts for each different gate type in $G$, the set of unique gate types. For each attempt, we first check if the generated response results in a valid circuit, i.e., the syntax adheres to the generic format of Boolean functions in which the input circuit is provided. If the response does not pass this check, we provide the LLM feedback about its incorrect format and ask it to try again. On the other hand, if the response passes this check,
we then check if the generated circuit follows our instructions
regarding the allowed set of Boolean operators (i.e., the allowed operators specified in the prompt instruction). If the circuit fails this second check, we provide the LLM feedback about the use of operators that are not allowed and ask it to try again.
However, if the circuit passes this second check, we move on to the third check where we evaluate the functional equivalence of the generated circuit to the original circuit. Again, if the generated circuit is not functionally-equivalent to the original circuit, we provide the LLM feedback about the non-equivalence and ask it to try again. Whereas, if the generated circuit passes this third check, we save the generated circuit (for later use in pirating circuits) and move on to the next gate type. Note that, for each gate type, if any of the three checks fails, we count it as a failed attempt (and increment the counter for the number of attempts), so each LLM has $M$ attempts to pass all three checks combined. As evidenced by our results, such interactive feedback-guided approach significantly improves LLMs performance (see Sec.~\ref{sec:num_attempts_and_improvement}).

The updated mathematical representation for this formulation is as follows: 
The final response (after potentially up to $M$ attempts) $R_i^M$ for the $i^{\text{th}}$ unique gate type is obtained from the underlying distribution 
\[
p(R_i^j| \mathcal{C}(R_i^{j-1}) \oplus \mathcal{T}(Q_i^j),\theta), \quad j\in\{1,2,\ldots,M\}, \quad \forall i \in \{1,2,\ldots,|G|\}, \tag{4}\label{eq:solution3}
\]
where $R_i^j$ denotes the LLM's response in the $j^{\text{th}}$ attempt for the $i^{\text{th}}$ unique gate type, and $\mathcal{C}(\cdot)$ denotes a function that analyzes the response for the three checks mentioned above (syntax, allowed operators, and functionality) and produces a feedback according to the result of the checks. Additionally, $\oplus$ denotes concatenation, which combines the feedback with the original circuit, creating the query for the next attempt. We use this final formulation to generate functionally-equivalent but structurally different circuits for all unique gate types in the target netlist.

Next, we describe our end-to-end flow
of pirating Verilog netlists, which includes \ccfilledgreen{A}  prompt Syntax translation For hardware netlists, \ccfilledgreen{B} pre-characterization and divide-and-conquer, and \ccfilledgreen{C} a feedback-guided interactive approach.

\subsection{Putting It All Together}\label{sec:putting_it_all_together}

Figure~\ref{fig:final_flow} illustrates the end-to-end flow. Given a netlist (or a set of netlists) to be pirated, we first perform pre-characterization (Solution \ccfilledgreen{B}), which analyzes the netlist(s) to extract the different gate types. Then, for each different gate type, we use the list of allowed Boolean operators in Table~\ref{tab:allowed_boolean_operators} to create prompts following the generic Boolean operator syntax (Solution \ccfilledgreen{A}). For instance, for each of the different \texttt{AND} gate types in the target netlist(s) (e.g., 2-input \texttt{AND} gate, 3-input \texttt{AND} gate, etc.), we create three prompts, one for each of the allowed Boolean operators: [\texttt{NAND}], [\texttt{NOR}], [\texttt{OR}, \texttt{NOT}]. This way, we create prompts for all different gate types for each of the corresponding allowed Boolean operators. Then, we pick a target LLM and query it for responses to these prompts, one after another. Additionally, as explained in Solution \ccfilledgreen{C}, after each response, we perform a series of checks (for syntax, use of only allowed Boolean operators, and functional equivalence). If any of these checks fail, we provide appropriate feedback to the LLM using a follow-up prompt. For instance, if the generated circuit fails the functionality check, we provide the following feedback ``\texttt{This is not correct because the functionality is not the same as the original circuit. Can you try again? Below is the original circuit:}'', followed by the original circuit in the generic Boolean function format provided in the initial prompt. In this way, we provide the LLM $M$ attempts to generate a circuit that passes all three checks. If, during any attempt, the LLM is successful, we save the generated circuit as a valid transformation of the original circuit so we can later use it for pirating netlists. For example Boolean transformations, see Sec.~\ref{sec:appendix_example_transformations} of the Appendix. On the other hand, even after $M$ attempts, if the LLM is unable to generate a a circuit that passes all three checks, we exit the loop and move on to the next allowed Boolean operator or to the next gate type. Thus, at the end, we obtain a dictionary of functionally equivalent transformations for all (or some, depending on the success of the LLM) different gate types using all (or some) of the different allowed Boolean operators for the corresponding gate type. For further information regarding the contribution of each solution within the framework, see the ablation study in Sec.~\ref{sec:appendix_ablation_study} of the Appendix.

Next, we describe how to pirate a given netlist using this dictionary of transformations. Recall that, for each different gate type, we have multiple transformations in the dictionary. In order to select the exact transformation to apply for a given gate when pirating a netlist, we devise five \textit{mapping strategies}: \textit{AND\_NOT}, \textit{NAND}, \textit{NOR}, \textit{OR\_NOT}, and \textit{random}. The \textit{NAND} mapping strategy only uses the [\texttt{NAND}] transformation, the \textit{AND\_NOT} mapping strategy only uses the [\texttt{AND}, \texttt{NOT}] transformation, and so on.\footnote{Since it is possible that some transformations might not exist for a given gate type, in that case, we pick a random transformation available in the dictionary for that gate type.} Finally, we pirate a given netlist using each of the five mapping strategies (one by one) by replacing the gates in the original netlist according to the transformation determined by the mapping strategy. Additionally, to overcome randomness, we repeat this process $N$ times and evaluate each of the $N\times5$ pirated versions using the piracy detection tools to obtain the similarity scores.

Note that, to ensure ease-of-use and wide application, the entire \myname{} flow described above is automated end-to-end, from characterizing netlists, to creating prompts, querying LLMs, performing the three checks, providing feedback to the LLMs, creating pirated versions of netlists, and finally evaluating them using the detection tools. Additionally, we ensure that the pirated netlists are functionally equivalent to the original netlists through exhaustive simulation-based testing. We further validate equivalence using \textit{Cadence Conformal Equivalence Checker}~\cite{cadence_lec}, an industry-standard commercial formal equivalence checker. We describe this in more detail in Sec.~\ref{sec:appendix_equivalence_of_pirated_netlists} of the Appendix.
Next, we demonstrate \myname{}'s efficacy in pirating netlists and evading a variety of detection tools.
\section{Results}\label{sec:results}

We conduct a detailed experimental investigation of the capabilities of different LLMs to pirate hardware IPs. Next, we detail our experimental setup.

\begin{table*}[t]
\centering
\caption{Details of detection tools used in our evaluation.}
\resizebox{\textwidth}{!}{
\begin{tabular}{ccccc}
\toprule
\textbf{Detection Tool} & GNN4IP~\cite{GNN4IP} & MOSS~\cite{moss_website} & Jplag~\cite{Jplag_github} & SIM~\cite{SIM_website} \\
\cmidrule{1-5}
\textbf{Source} & \begin{tabular}[c]{@{}c@{}} GNN4IP Repository~\cite{hw2vec_github} \end{tabular} & Internet Submission Method~\cite{moss_website} & Jplag Repository~\cite{Jplag_github} & \begin{tabular}[c]{@{}c@{}}Source Code~\cite{SIM_website}\end{tabular} \\
\cmidrule{1-5}
\textbf{\begin{tabular}[c]{@{}c@{}}Similarity Scores Range\end{tabular}} & [-1,1] & [0,1] & [0,1] & [0,1]\\ \cmidrule{1-5}
\textbf{\begin{tabular}[c]{@{}c@{}}Detection Threshold\end{tabular}} & 0 & 0.2 & 0.3 & 0.3\\ \cmidrule{1-5}
\textbf{\begin{tabular}[c]{@{}c@{}}Notes\end{tabular}} & \begin{tabular}[c]{@{}c@{}}Designed for Verilog,\\Most accurate tool\\for our case\end{tabular} & \begin{tabular}[c]{@{}c@{}}Supports Verilog,\\Stricter threshold as it was primarily\\designed for software code\end{tabular} & \begin{tabular}[c]{@{}c@{}}Doesn't support Verilog,\\Relatively strict threshold\\as we use it in text mode\end{tabular} & \begin{tabular}[c]{@{}c@{}}Doesn't support Verilog,\\Relatively strict threshold\\as we use it in text mode\end{tabular}\\
\bottomrule
\end{tabular}
}
\label{tab:detection_tools_table}
\end{table*}

\begin{figure*}[t]
    \centering
    \includegraphics[width=\textwidth,trim={0.2cm 0.2cm 0.2cm 0.2cm},clip]{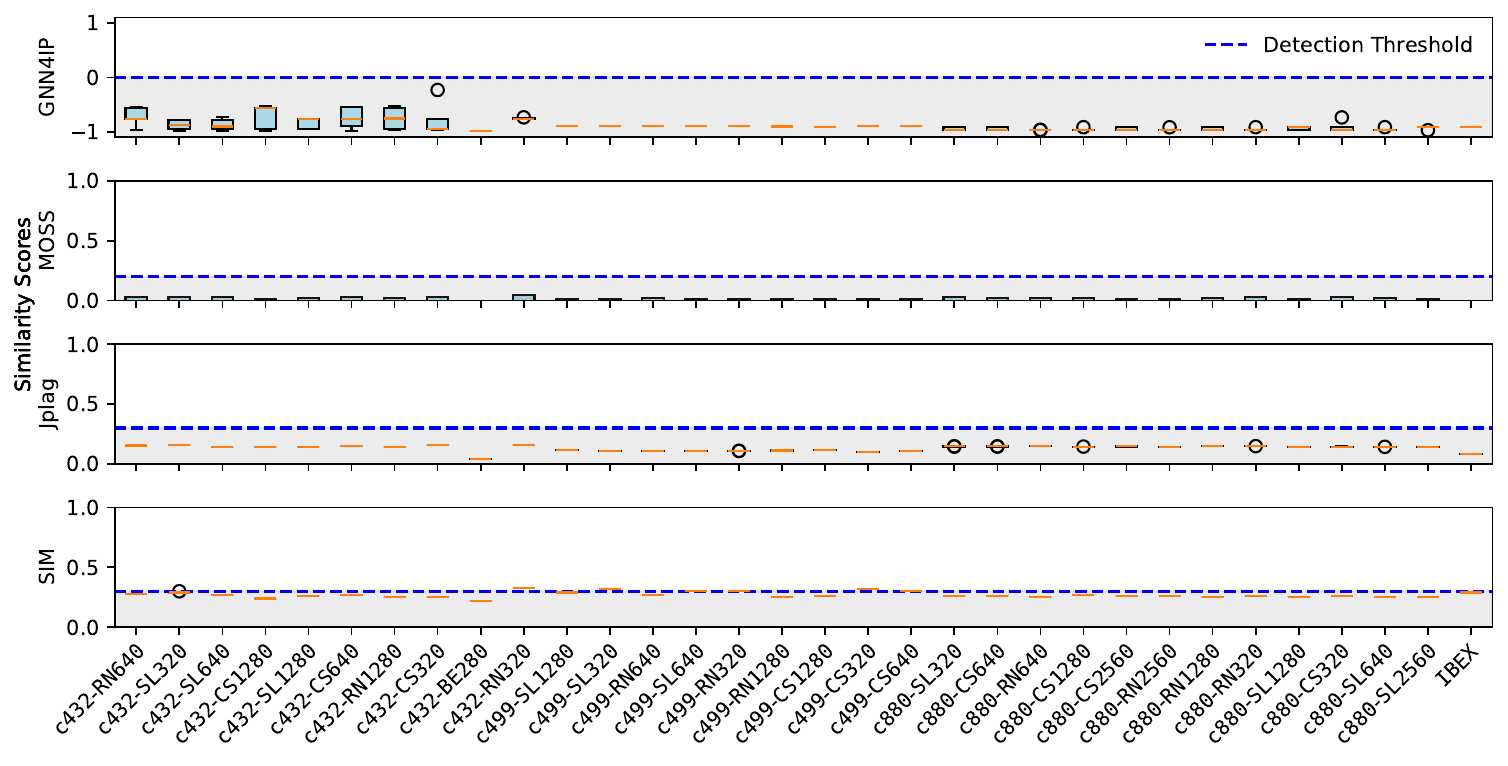}
    \caption{\myname{}'s best performance against GNN4IP~\cite{GNN4IP}, MOSS~\cite{moss_website}, Jplag~\cite{Jplag_github}, and SIM~\cite{SIM_website}.}
    \label{fig:main_boxplot}
\end{figure*}

\subsection{Experimental Setup}\label{sec:experimental_setup}

We implement \myname{} using \textit{Python}. We set $M$, the maximum number of attempts available to the LLMs, to be 5. We set $N$, the number of pirated netlists created for each mapping strategy to capture the effect of randomness (Sec.~\ref{sec:putting_it_all_together}), to be 5. We use a dataset of 31 different Verilog netlists from the GNN4IP repository for our experiments~\cite{hw2vec_github}. We choose these netlists because of two reasons: (i)~GNN4IP is trained on them, making this a more difficult setting for our attack than the typical setting where one pirates netlists that the detection tool has not seen before, e.g., netlists from the testing set of GNN4IP. We adhere to this challenging scenario to highlight \myname{}'s remarkable proficiency in effectively pirating netlists. (ii)~Additionally, we are constrained by these available netlists in GNN4IP to perform a fair evaluation of GNN4IP. However, these netlists are small, and small netlists are difficult to pirate since there is a limited set of gates to work with and detection tools perform very well on them. However, to showcase \myname{}'s scalability and ramifications, we also test it on large netlists, \texttt{IBEX}~\cite{IBEX} and \texttt{MOR1KX}~\cite{MOR1KX} processors, and a \texttt{GPS}~\cite{GPS} module. We chose these netlists because of their significant design complexity, practical relevance, and widespread adoption across various applications. Our target netlists range from a few hundred gates to hundreds of thousands of gates. We provide the netlist size metrics in Table~\ref{tab:design_sizes} in Sec.~\ref{sec:appendix_design_sizes} of the Appendix.

\begin{figure*}[t]
    \centering
    \includegraphics[width=\textwidth,trim={0.2cm 0.2cm 0.2cm 0.2cm},clip]{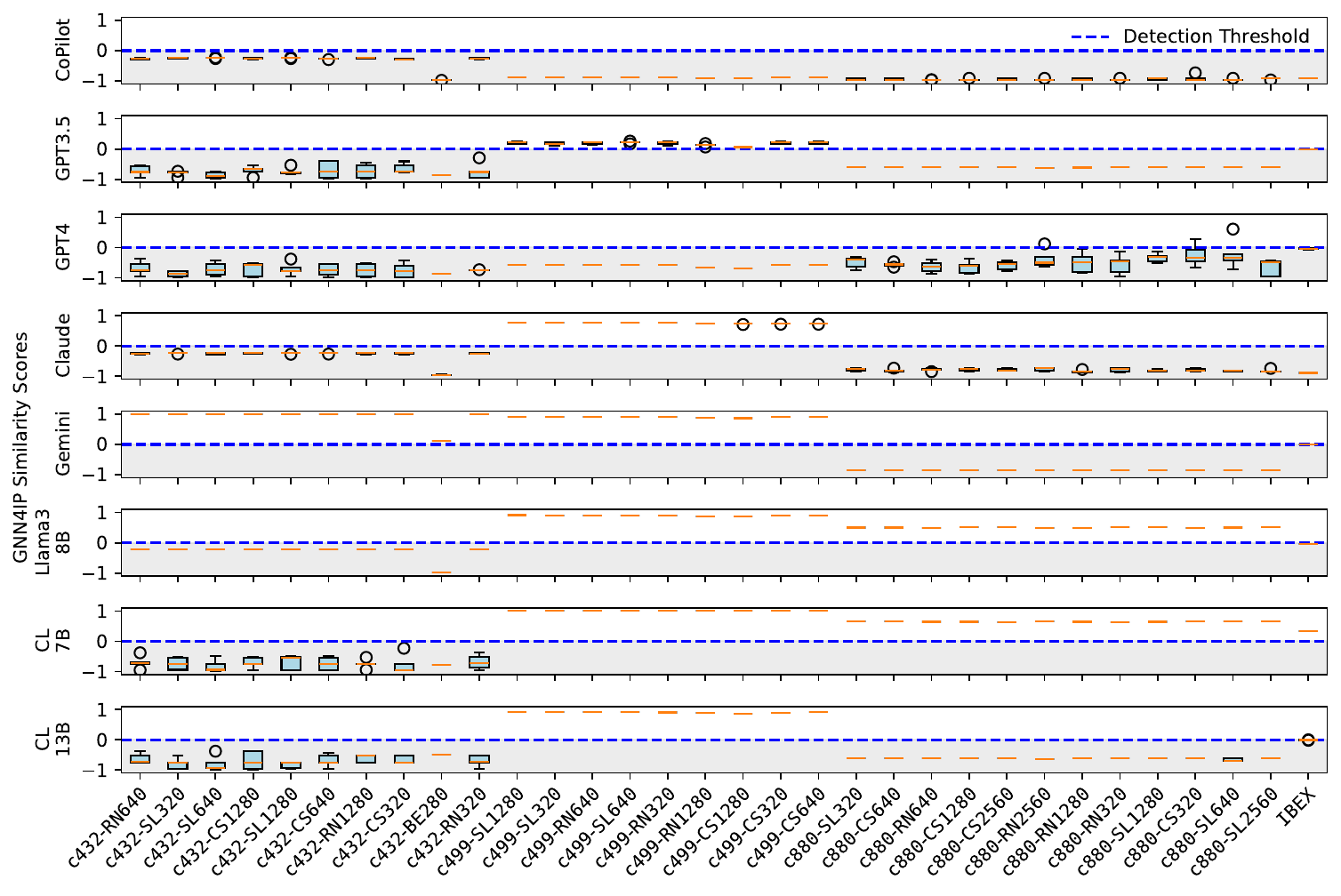}
    \caption{Distribution of GNN4IP~\cite{GNN4IP} similarity scores for different LLMs in \myname{}'s framework.}
    \label{fig:GNN4IP_boxplot}
\end{figure*}

\noindent\textbf{LLMs.} For a thorough analysis, we select eight representative LLMs for our evaluations:
\begin{itemize}[leftmargin=*]
    \item \textbf{CoPilot} from Microsoft uses the Prometheus model, and iteratively generates search queries, to combine Bing search results with OpenAI's GPT-4 and GPT-4 Turbo LLMs to produce responses~\cite{how_copilot_works}. We use CoPilot via \url{http://copilot.microsoft.com}.
    \item \textbf{GPT-3.5} models from OpenAI that can understand and generate natural language or code and have been optimized for chat
    and instruction based tasks~\cite{openai_gpt3dot5_description}. We use the \texttt{gpt-3.5-turbo-16k} model in our experiments.
    \item \textbf{GPT-4} models improve on GPT-3.5 and can understand as well as generate natural language or code with greater accuracy than any of the previous GPT models~\cite{openai_gpt4_description}. Also, GPT-4 is one of the most advanced general-purpose LLMs available today. We use the \texttt{gpt-4-turbo} model, the most advanced GPT-4 model, in our experiments.
    \item \textbf{Claude} models from Anthropic can perform complex analysis, tasks with multiple steps, and higher-order math and coding tasks~\cite{claude3_description}. Also, they have low hallucination rates~\cite{claude3_description}. We use the \texttt{claude-3-opus-20240229} model, the most advanced Claude model, in our experiments.
    \item \textbf{Gemini} models from Google are built
    for reasoning
    across text, images, audio, video, and code~\cite{gemini1dot5_website}. Gemini 1.0 was the first model to outperform human experts on the Massive Multitask Language Understanding benchmark~\cite{gemini1_website}. We use the \texttt{gemini-1.0-pro} model in our experiments.
    \item \textbf{Llama2} is a set of open-source LLMs developed by Meta. At the time of release, Llama 2 outperformed the other open-source models across all benchmarks~\cite{llama2_website}. CodeLlama is a code-specialized version of Llama2 that was created by further training Llama2 on its code-specific datasets~\cite{codellama_announcement}. In our experiments we use two variants of CodeLlama: \texttt{CodeLlama-7b-Instruct-hf} and \texttt{CodeLlama-13b-Instruct-hf} (denoted as CL-7B and CL-13B in our evaluations) from HuggingFace~\cite{codellama_huggingface}.
    \item \textbf{Llama3} is a recently released and highly capable openly available set of LLMs~\cite{llama3_website}. These models have greatly improved capabilities like reasoning, code generation, and instruction following. We use the \texttt{Meta-Llama-3-8B-Instruct} (denoted as Llama3-8B) model in our experiments.
\end{itemize}
Our selection of LLMs represents some of the most advanced LLMs available today from a variety of organizations, as well as current state-of-the-art and recently released publicly available LLMs capable of performing code-related tasks.

\noindent\textbf{Detection Tools.} To evaluate the success of these LLMs in pirating Verilog netlists, we use the piracy/similarity measurement tools described in Table~\ref{tab:detection_tools_table}.
As explained in Sec.~\ref{sec:background_piracy_detection}, our selection consists of highly accurate, widely-used, and popular techniques ranging over multiple decades. The remainder of the section is organized as follows: First, we provide the main piracy results against these detection tools (Sec.~\ref{sec:results_main_result}). Then, we provide further results and analysis of the performance of the LLMs against each detection tool separately (Secs.~\ref{sec:results_against_GNN4IP}-\ref{sec:results_against_SIM}). We also perform more analyses of the different mapping strategies and the number of attempts available to the LLMs (Secs.~\ref{sec:analysis_of_mapping_strategies_GNN4IP},~\ref{sec:num_attempts_and_improvement}). Then, we demonstrate the ramifications of \myname{} through case studies on practical netlists, \texttt{IBEX} and \texttt{MOR1KX} processors and a \texttt{GPS} module (Secs.~\ref{sec:results_case_study},~\ref{sec:results_case_study_scalability}).
Finally, we summarize the key characteristics of LLMs in an effort to understand their performances in Sec.~\ref{sec:results_llms_characteristics}.

\begin{figure*}[t]
    \centering
    \includegraphics[width=\textwidth,trim={0.2cm 0.2cm 0.2cm 0.2cm},clip]{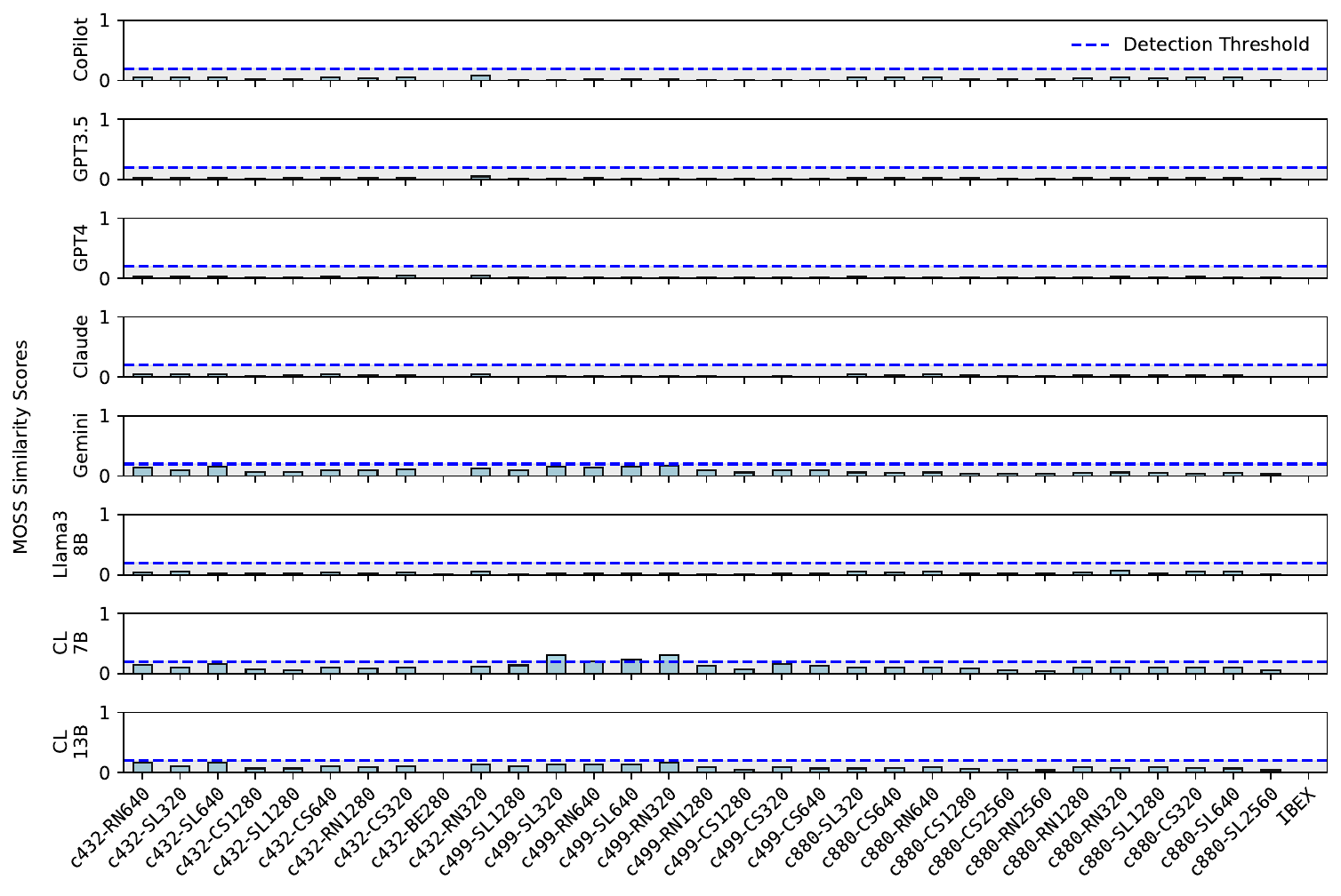}
    \caption{MOSS~\cite{moss_website} similarity scores for different LLMs in \myname{}'s framework.}
    \label{fig:MOSS_boxplot}
\end{figure*}

\subsection{Main Piracy Results}\label{sec:results_main_result}
Figure~\ref{fig:main_boxplot} shows the distribution of best (i.e., the lowest) similarity scores from the four detection tools for the 32 netlists in our dataset.\footnote{Note that, to showcase the best results achieved using \myname{}, the similarity scores plotted are the best (i.e., the lowest) scores over all mapping strategies and LLMs. We provide more fine-grained results of the performance of different LLMs and mapping strategies in the subsequent subsections.} It is clear that using \myname{}, we are successfully able to pirate all 32 netlists against all four detection tools with very limited variance in performance. Note that since MOSS limits use to 100 queries per day per user~\cite{moss_website}, we randomly picked one of the $N=5$ pirated netlists for each mapping strategy and queried MOSS for similarity. Hence, the plots for MOSS are bar plots showing the single similarity scores instead of box plots for the distribution of similarity scores. Also note that the Jplag and SIM similarity scores are higher (compared to MOSS) because Verilog netlists have keywords (e.g., \texttt{and}, \texttt{nand}, etc.) that are repeated frequently, and since these detection tools are used in text mode, such repeated keywords contribute to high similarity scores. Nonetheless, \myname{} successfully 
pirates all netlists. 
Additionally, due to our divide-and-conquer approach and saving of generated transformations, the runtime of \myname{} for any given LLM is in the order of a few minutes. Furthermore, the performance overheads of our pirated netlists are also reasonable (see Sec.~\ref{sec:appendix_performance_overheads} in the Appendix).
\subsection{LLMs Against GNN4IP}\label{sec:results_against_GNN4IP}

To evaluate the LLMs' ability to pirate Verilog netlists, we first analyze GNN4IP's similarity scores between each of the pirated netlists and the original netlists. Figure~\ref{fig:GNN4IP_boxplot}, summarizes these values across all 32 netlists for each LLM.
Note that the distribution of similarity scores plotted for each netlist for each LLM are for the best mapping strategies for that netlist and LLM.

Here are the key takeaways from the figure: (i)~Most LLMs (CoPilot, GPT-3.5, GPT-4, Claude, and CL-13B, i.e., CodeLlama-13B) are successfully able to evade GNN4IP detection for the majority of the netlists. (ii)~Some LLMs (Gemini, Llama3-8B, and CL-7B) are unable to pirate the majority of netlists (the reason behind this in explained in Sec.~\ref{sec:num_attempts_and_improvement}). The worst performing LLMs are CL-7B and Llama3-8B, only evading detection on 10 and 11 of the 32 netlists, respectively. (iii)~CL-13B performed significantly better (23 successes) than its smaller version, CL-7B (10 successful netlists).
(iv)~Overall, GPT-4 and CoPilot (which uses GPT-4 internally) achieve the best performance, i.e., lowest GNN4IP similarity scores.

\subsection{LLMs Against MOSS}\label{sec:results_against_MOSS}
Here, we repeat the evaluation process using MOSS as the piracy detection tool.
As explained in Sec.~\ref{sec:results_main_result}, due to restrictions on the number of queries, the plots for MOSS in Figure~\ref{fig:MOSS_boxplot} are bar plots showing the single similarity score instead of box plots showing the distribution of similarity scores. Note that these single similarity scores still provide enough information to analyze the performance of LLMs against MOSS. Also, as in Sec.~\ref{sec:results_against_GNN4IP}, the similarity score plotted for each netlist for each LLM is for the best mapping strategy for that netlist and LLM.

Here are the key takeaways: (i)~All LLMs except CL-7B evade MOSS for all netlists. (ii)~As with GNN4IP, CL-13B performs better than the smaller CL-7B. (iii)~Notably, Llama3-8B is at par with larger models (e.g., GPT-3.5 and GPT-4). 

\subsection{LLMs Against JPlag}\label{sec:results_against_Jplag}
Here, we use the same evaluation procedure using Jplag as the piracy detection tool. In the interest of space, we plot the figure in Sec.~\ref{sec:appendix_llms_against_Jplag_SIM} of the Appendix (Figure~\ref{fig:JPlag_boxplot}) and explain the key takeaways here: (i)~Most closed-source LLMs (CoPilot, GPT-3.5, GPT-4, Claude) successfully evade detection across all 32 netlists. (ii)~As seen before, CL-13B performs better than the smaller CL-7B. (iii)~As with MOSS, the open-source Llama3-8B performs almost as well as the larger models, successfully bypassing JPlag for all but one netlists.

\subsection{LLMs Against SIM}\label{sec:results_against_SIM}
The evaluation is again repeated using SIM as the piracy detector, with Figure~\ref{fig:Sim_boxplot3} (in Appendix Sec.~\ref{sec:appendix_llms_against_Jplag_SIM}) illustrating the similarity scores. Here are the key takeaways: (i)~Due to the lack of compatibility with Verilog and the text mode of operation of SIM, it results in unusually high similarity scores because Verilog keywords (e.g., \texttt{nand}, \texttt{and}, etc.) are repeated frequently in the netlists. (ii)~Nonetheless, GPT-3.5 and GPT-4 still evade SIM for 25 netlists. (iii)~Llama3-8B, with 11 successes, performs the best among open-source models.

\begin{myframe}
\textbf{Finding 1.} Overall, GPT-4 and CoPilot achieve the best performance in successfully pirating netlists against all four detection tools.
\end{myframe}

\begin{myframe}
\textbf{Finding 2.} Overall, CodeLlama-13B performs significantly better than the smaller CodeLlama-7B. More often than not, Llama3-8B performs better (for our task) than the CodeLlama models (which are based on Llama2).
\end{myframe}

\subsection{Analysis of Mapping Strategies}\label{sec:analysis_of_mapping_strategies_GNN4IP}
\begin{figure}[t]
    \centering
    \includegraphics[width=0.48\textwidth,trim={0.2cm 0.2cm 0.2cm 0.0cm},clip]{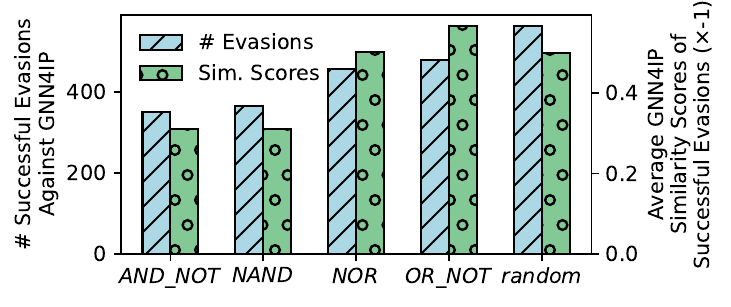}
    \caption{Performance of mapping strategies against GNN4IP.}
    \label{fig:mapping_strat_analysis_GNN4IP}
\end{figure}

So far, we analyzed the main piracy results against four detection tools, and the performance of different LLMs against different tools. Now, we take a closer look at the performance of the five mapping strategies, \textit{AND\_NOT}, \textit{NAND}, \textit{NOR}, \textit{OR\_NOT}, and \textit{random}. More specifically, to understand the relative performance of these mapping strategies, we analyze them in terms of the number of successful instances of evasions (over all netlists and all LLMs) and the average similarity scores of those instances against GNN4IP (Figure~\ref{fig:mapping_strat_analysis_GNN4IP}). It is evident that the \textit{random} strategy yields the largest number of successful evasions. This makes intuitive sense because, with the \textit{random} strategy, a given gate can be replaced with any of its transformations, leading to different structures in the pirated netlist, whereas, with other strategies, the pirated netlist is likely to have similar structures due to the possibility of more deterministic replacements. This is also reflected in the low (note that the second y-axis is inverted, i.e., multiplied by -1) average GNN4IP similarity scores compared to most other strategies. We observe similar results against MOSS, Jplag, and SIM (see Sec.~\ref{sec:appendix_analysis_of_mapping_strategies_MOSS_Jplag_SIM} in the Appendix).

\begin{myframe}
\textbf{Finding 3.} All five mapping strategies result in successful pirated netlists against all detection tools, with the \textit{random} mapping strategy showing the best performance in terms of the similarity scores. 
\end{myframe}

\subsection{LLMs' Performance Comparison}\label{sec:num_attempts_and_improvement}
\begin{figure}[t]
    \centering
    \includegraphics[width=0.48\textwidth,trim={0.2cm 0.2cm 0.2cm 0.0cm},clip]{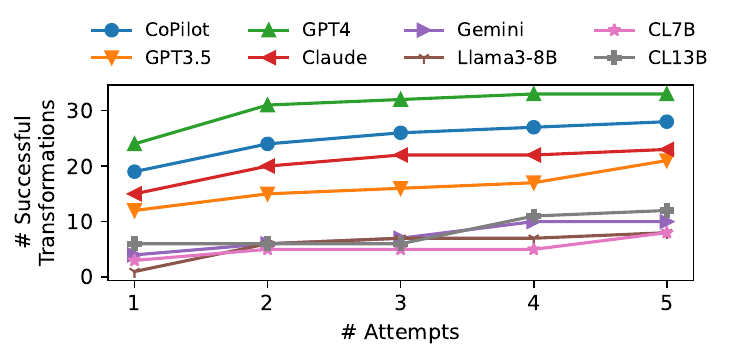}
    \caption{Comparison of number of attempts for successful transformations using different LLMs.}
    \label{fig:num_attempts}
\end{figure}

Next, we compare the performance of the LLMs in generating successful transformations according to the allowed Boolean operators in Table~\ref{tab:allowed_boolean_operators} (e.g., \texttt{AND} gate using \texttt{NOR} operators, etc.). Recall that we allow each LLM a maximum of $M=5$ attempts for each different gate type. Additionally, if an attempt fails, we also provide fine-grained feedback about syntax, use of correct Boolean operators, or functionality in allow the LLM to fix its mistakes. To that end, Figure~\ref{fig:num_attempts} shows the total number of successful transformations generate by different LLMs as a function of the number of attempts. It is evident that, after 5 attempts, there are two classes of LLM in terms of number of successful transformations. The first class consists of GPT-4, CoPilot, Claude, and GPT-3.5, with 33, 28, 23, and 21 successful transformations, respectively. The second class consists of CL-13B, Gemini, 
Llama3-8B, and CL-7B, with significantly fewer successful transformations. This explains why the LLMs from the second class are sometimes unable to evade some
detection tools. A surprising observation is that Gemini (one of the closed-source LLM that performs similar to the GPT models on other common tasks) struggles with our task of rewriting circuits. The exact reason behind this is difficult to know, however, a possible reason could be a lack of enough Verilog/circuit training data. Another observation from the figure is that all LLMs improve with more attempts and feedback. This validates our Solution \ccfilledgreen{C} of devising a feedback-guided interactive formulation for our task. 

\begin{figure}[t]
    \centering
    \includegraphics[width=0.48\textwidth,trim={0.2cm 0.2cm 0.2cm 0.0cm},clip]{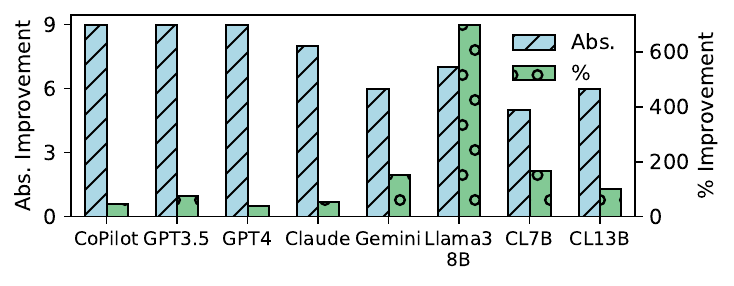}
    \caption{Absolute and percentage improvements (attempt 5 vs. attempt 1) in successful transformations.}
    \label{fig:num_attempts_abs_and_percent_impr}
\end{figure}

We also analyze the impact of multiple attempts through the absolute and percentage improvements (attempt 5 vs. attempt 1) in the number of successful transformations in Figure~\ref{fig:num_attempts_abs_and_percent_impr}. Overall, all LLMs benefit from the multiple attempts, with larger and more capable LLMs (CoPilot, GPT-3.5, GPT-4, and Claude) showing the most absolute improvement, i.e., most improvement in number of successful mappings at attempt 5 compared to attempt 1. However, smaller LLMs (such as Llama3-8B, CL-7B, and CL-13B), especially Llama3-8B, really leverage the multiple attempts with feedback and improve drastically (e.g., 700\% improvement in Llama3-8B) over their poor performance in the first attempt.

\begin{myframe}
\textbf{Finding 4.} We find two classes of LLMs in terms of the number of successful transformations they generate, explaining why LLMs from the second class are sometimes unable to evade detection.
\end{myframe}

\begin{myframe}
\textbf{Finding 5.} Our interactive formulation with multiple attempts and fine-grained feedback improves the performance of all LLMs, especially smaller, less capable LLMs, such as Llama3-8B.
\end{myframe}

\subsection{Case Study on the \texttt{IBEX} Processor: Ramifications of \myname{}}\label{sec:results_case_study}
\begin{figure}[t]
    \centering
    \includegraphics[width=0.48\textwidth,trim={0.2cm 0.2cm 0.2cm 0.0cm},clip]{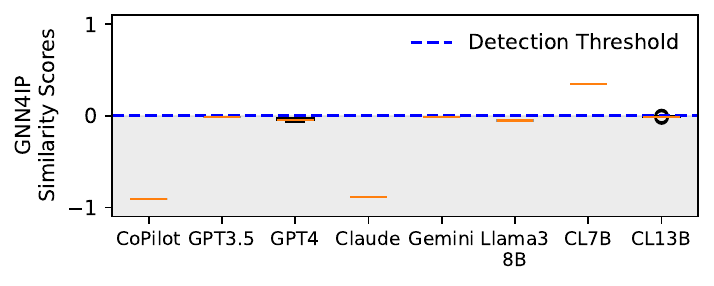}
    \caption{Comparison of GNN4IP similarity scores for \texttt{IBEX} processor pirated using different LLMs.}
    \label{fig:case_study_LLM_sim_scores_IBEX}
\end{figure}

In this subsection, we demonstrate and discuss the performance of \myname{} on a real-world netlist, the 
\texttt{IBEX} processor~\cite{IBEX} in more detail. Specifically, we run our end-to-end automated flow of \myname{} to pirate the processor using our mapping strategies and the corresponding transformations obtained from the eight LLMs. Then, we query GNN4IP to get the similarity scores between our pirated netlists and the original netlist. Note that, similar to related works, we assume full-scan access to ensure compatibility of the netlists with GNN4IP~\cite{gohil2024attackgnn}. Figure~\ref{fig:case_study_LLM_sim_scores_IBEX} compares the distribution of the GNN4IP similarity scores for the eight LLMs. Note that, as earlier, the distributions are for the best-performing strategy. We observe that seven out of the eight LLMs (all except CL-7B) successfully evade GNN4IP. 
Thus, \myname{} easily fools GNN4IP into classifying pirated versions of \texttt{IBEX} as not pirated. Moreover, the distribution of the GNN4IP similarity scores is extremely low for CoPilot and Claude, meaning that not only is GNN4IP evaded, the magnitude of the incorrect detection is extremely high. This case study demonstrates the capabilities of \myname{}, which can lead to piracy of practical netlists, and failure of the state-of-the-art piracy detection tool in catching it.

\subsection{Case Study on Larger Netlists: Scalability of \myname{}}\label{sec:results_case_study_scalability}
Recall that \myname{} first generates and caches (i.e., saves) valid gate transformations. Then, for any given target netlist, it creates a pirated netlist gate-by-gate. So, working with a larger netlist will only increase the runtime linearly with the number of gates, and thus will be easily manageable. We validate this scalability of \myname{} by further experimenting with even larger netlists, containing hundreds of thousands of gates. More specifically, we target two open-source netlists: a \texttt{GPS} module from Common Evaluation Platform~\cite{GPS}, containing $\approx193$K gates, and an \texttt{MOR1KX} processor~\cite{MOR1KX}, containing $\approx158$K gates. We observed that \myname{} generates pirated netlists within seconds. Additionally, these netlists successfully evade MOSS~\cite{moss_website}, Jplag~\cite{Jplag_github}, and SIM~\cite{SIM_website}, but surprisingly, GNN4IP~\cite{GNN4IP} always classifies these \myname{}-generated netlists as pirated. This unusual result might lead one to believe that GNN4IP thwarts \myname{} for larger netlists. However, a closer evaluation reveals a surprising observation. To validate the efficacy of GNN4IP for these large netlists, we query GNN4IP to predict the similarity between the original \texttt{GPS} and \texttt{MOR1KX} netlists, and observe that GNN4IP yields an extremely high similarity score of $0.97$. This means that GNN4IP has a high bias towards classifying large netlists as pirated,
which explains why \myname{}-generated netlists are unable to evade GNN4IP.
This bias of GNN4IP 
makes it unsuitable for evaluating such netlists.

\begin{myframe}
\textbf{Finding 6.} GNN4IP, the current state-of-the-art hardware IP piracy detector, struggles against \myname{} for large netlists.
\end{myframe}

\subsection{LLMs' Characteristics}\label{sec:results_llms_characteristics}
In the previous sections, we evaluated the performance of \myname{} in evading detection tools. Here, we delve deeper and provide some insights about
the characteristics of LLMs that make some LLMs perform better than others in the context of hardware IP piracy.
\begin{itemize}[leftmargin=*]
    \item Model Size Matters: Large LLMs, with potentially trillions of parameters, perform best.
    \item Training Data Size Matters: Latest version of Llama (Llama3-8B) outperforms the older Llama2 models, again, potentially due to its $>7\times$ training data size~\cite{llama3_blog,llama2_website}.
    \item Open vs. Closed Source LLMs: Closed source LLMs still have a fairly decent margin compared to open source LLMs for our task.
    \item Potential of Smaller LLMs: With proper feedback and multiple attempts, smaller LLMs correct their mistakes.
\end{itemize}

\begin{myframe}
\textbf{Summary of Results.} \myname{} successfully pirates all given netlists (including the \texttt{IBEX} and \texttt{MOR1KX} processors and the \texttt{GPS} module) 
and evades all detection tools with only minutes of runtime cost.
\end{myframe}

\section{Related Work and Discussion}\label{sec:related_work}

Here, we first discuss the need for \myname{} over simply using Verilog-fine-tuned LLMs. Then we discuss related works on hardware and software intellectual property (IP) piracy and how our work is different from them. We also discuss applicability to other detection tools and potential countermeasures against \myname{}. We also discuss the impact of netlist obfuscation in Sec.~\ref{sec:appendix_discussion_on_designs_protected_by_obfuscation} of the Appendix.

\subsection{Evaluating Verilog-fine-tuned LLMs}
As observed in Challenge \ccfilledred{1} in Sec.~\ref{sec:methodology_pirating_firm_ips}, various general-purpose open-source and closed-source LLMs struggle in understanding and rewriting simple netlist written in Verilog hardware description language. So, a natural question could be about the possibility of using LLMs fine-tuned on Verilog data instead of using the solutions we described. To address this, first, note that fine-tuning LLMs is computationally much more expensive than our current approach. Nonetheless, we tested VeriGen (an LLM fine-tuned on a large corpus of Verilog dataset)~\cite{thakur2023verigen} by asking it to rewrite the simple gate-level netlist shown in Listing~\ref{listing:prelim_formulation_prompt}. To account for the LLMs' non-deterministic responses,
we repeated the experiment $10$ times, but VeriGen was unsuccessful in either generating a syntactically correct code, i.e., a valid gate-level netlist, or a functionally-equivalent netlist even once out of the $10$ times. This demonstrates that even fine-tuned LLMs struggle in the context of our task of rewriting gate-level netlists and emphasize the need for \myname{} to successfully pirate netlists.

\subsection{Evading Hardware IP Piracy Detection}\label{sec:discussion_evading_hardware_IP_piracy_detection}
Various works have demonstrated strategies to evade hardware IP detection tools, primarily targeting GNN4IP~\cite{gohil2024attackgnn, alrahis2023poisonedgnn}. \PoisonedGNN{}, exploits the susceptibility of graph neural networks (GNNs) to data poisoning attacks by injecting backdoor triggers at the register and gate-level of circuit designs within GNN4IP's training dataset~\cite{alrahis2023poisonedgnn}. The resulting accuracy of the target GNN model is then reduced, enabling pirated circuits to successfully evade GNN4IP detection. 
However, unlike \PoisonedGNN{}, our work (i)~does not require access to the GNN's training procedure, (ii)~evades detection without changing the GNN's parameters or training dataset, (iii)~assumes only black-box access to the target GNN, and (iv)~does not design backdoors specific to target detection tools, but evades multiple types of detection tools, including GNN4IP.

\AttackGNN{}, another recent technique, is perhaps the closest to our work in terms of evading GNN4IP~\cite{gohil2024attackgnn}.
By training a reinforcement learning agent, \AttackGNN{} learns to perturb netlists that evade GNN4IP detection. However, there are critical differences in terms of methods, results, and impact. Unlike \AttackGNN{}, \myname{} (i)~lowers the technical expertise barrier: our work does not need detailed understanding of reinforcement learning methods; rather, it simply requires prompting off-the-shelf LLMs, (ii)~does not require time-consuming training procedure, and generates pirated netlists directly using LLMs assisted with a quick feedback-guided formulation, and (iii)~is not tailored to only generating adversarial examples that evade machine-learning-based detectors such as GNN4IP; rather, \myname{} evades a variety of detection tools.

\subsection{Evading Software IP Piracy Detection}
Previous works have investigated strategies in which software IP piracy detection tools can be successfully evaded. In particular, MOSSAD~\cite{mossad} details an automated program transformation algorithm that is able to successfully evade MOSS and JPlag. Through combining genetic programming techniques and domain-specific knowledge, MOSSAD is able to effectively generate multiple semantically equivalent variants of a given program which are evaluated as no more suspicious than a non-plagiarized counterpart~\cite{mossad}. However, a key component in MOSSAD is adding code lines that do not affect the final output. Using such an approach for hardware IPs is not feasible since unused gates are trivially optimized out in Verilog netlists. In fact, MOSSAD highlights this as a weakness in its approach since when working with compiled and optimized assembly code, MOSSAD fails.

LLMs have also been evaluated in their ability to generate software that bypasses plagiarism detection tools. Researchers determined that when utilized by students on programming assignments, GPT-J (a 6-billion parameter LLM) is able to generate functional code that evades MOSS detection~\cite{biderman2022fooling}. However, such techniques focus on simply generating new software programs (not pirating existing programs) using an LLM, which is clearly not feasible for hardware codes because of LLMs' terrible performance in understanding and generating netlists (as evidenced in Sec.~\ref{sec:methodology}).

\subsection{Other Detection Tools and Potential Countermeasures}\label{sec:discussion_other_detection_tools_and_countermeasures}
\noindent\textbf{Other Detection Tools.} Since a detection tool (other than the ones we evaluated) can be built on different principles, it is difficult to guarantee \myname{}'s success against new tools. However, since \myname{} evades detection tools based on a wide variety of principles (e.g., GNN, winnowing, greedy string tiling, etc.), we are hopeful that similar results would hold for new detection tools too. Moreover, although the actual evasion performance could vary, the techniques we developed, i.e., prompt syntax translation, pre-characterization and divide-and-conquer, and different kinds of feedback, would still likely be helpful in improving piracy performance.

\noindent\textbf{Potential Countermeasures.} There can be a few different potential countermeasures against our work. (i)~Re-training models like GNN4IP with our pirated netlists included in the training set to increase the robustness of detection. However, research has shown limitations of such approaches for GNNs~\cite{gosch2023adversarial}. (ii)~Another potential countermeasure could be watermarking to identify LLM-generated text~\cite{kirchenbauer2023watermark}, however, since we don't directly use LLMs to generate pirated netlists, but process the output from LLMs to aid the piracy process, the applicability of watermarking as a countermeasure against our attack is unclear and needs further investigation.
\section{Conclusion}\label{sec:conclusion}
Large language models have become increasingly capable of understanding and generating code, leading to their adoption into the hardware design industry. However, we observe that these models also can be maliciously employed to attack vulnerabilities within this design process, particularly resulting in additional security concerns regarding hardware IP piracy.

To demonstrate this, we devised~\myname{}, a first-of-its-kind end-to-end automated, efficient, and practical framework that leverages the logical capabilities of LLMs to successfully pirate circuit designs in the form of Verilog netlists. 
Since LLMs are trained on very limited Verilog data, their off-the-shelf performance in pirating netlists is poor, so we formulate various solutions to achieve successful piracy. In particular, we perform syntax translation from netlists to generic Boolean function format, allowing the LLMs to better understand the circuit design.
We use pre-characterization and divide-and-conquer techniques to overcome context window limitations and ensure scalability to large netlists. We also incorporate a fine-grained feedback-guided iterative flow to mitigate error-prone responses, ensuring reliability.

Our experimental results confirm that \myname{} is able to evade detection against four state-of-the-art piracy detection tools across every tested netlist. We test on the netlists from the GNN4IP repository as those netlists are seen by GNN4IP during training and are relatively small in size, both factors making them difficult to pirate successfully. Despite this, overall, GPT-3.5 and GPT-4 demonstrate the best ability to pirate the netlists. We also observe that the smaller LLMs (Llama3-8B, CodeLlama-7B, CodeLlama-13B) derive the most relative improvement from our feedback-guided flow. Finally, we highlight the ramifications of our work through case studies on \texttt{IBEX} and \texttt{MOR1KX} processors and a \texttt{GPS} module,
demonstrating both, the capabilities of \myname{} and the limitations of current piracy detectors.
\section*{Acknowledgment}
The authors acknowledge the support from the Purdue Center
for Secure Microelectronics Ecosystem – CSME\#210205. This
work was also partially supported by the National Science
Foundation (NSF CNS–1822848 and NSF DGE–2039610).
\bibliographystyle{IEEEtranS}
\balance
\bibliography{main}

\clearpage
\section*{Appendix}\label{sec:appendix}
\setlength{\oldtabcolsep}{\tabcolsep}
\setlength{\tabcolsep}{1pt}
\begin{table*}[t]
\caption{Sizes of target netlists in terms of number of gates.}
\label{tab:design_sizes}
\resizebox{\textwidth}{!}{%
\begin{tabular}{cccccccccccccccccccccccccccccccccc}
\toprule
\rotatebox{75}{\textbf{\texttt{c432-RN640}}} & 
\rotatebox{75}{\textbf{\texttt{c432-SL320}}} & 
\rotatebox{75}{\textbf{\texttt{c432-SL640}}} & 
\rotatebox{75}{\textbf{\texttt{c432-CS1280}}} & 
\rotatebox{75}{\textbf{\texttt{c432-SL1280}}} & 
\rotatebox{75}{\textbf{\texttt{c432-CS640}}} & 
\rotatebox{75}{\textbf{\texttt{c432-RN1280}}} & 
\rotatebox{75}{\textbf{\texttt{c432-CS320}}} & 
\rotatebox{75}{\textbf{\texttt{c432-BE280}}} & 
\rotatebox{75}{\textbf{\texttt{c432-RN320}}} & 
\rotatebox{75}{\textbf{\texttt{c432-SL1280}}} & 
\rotatebox{75}{\textbf{\texttt{c499-SL320}}} & 
\rotatebox{75}{\textbf{\texttt{c499-RN640}}} & 
\rotatebox{75}{\textbf{\texttt{c499-SL640}}} & 
\rotatebox{75}{\textbf{\texttt{c499-RN320}}} & 
\rotatebox{75}{\textbf{\texttt{c499-RN1280}}} & 
\rotatebox{75}{\textbf{\texttt{c499-CS1280}}} & 
\rotatebox{75}{\textbf{\texttt{c499-CS320}}} & 
\rotatebox{75}{\textbf{\texttt{c499-CS640}}} & 
\rotatebox{75}{\textbf{\texttt{c880-SL320}}} & 
\rotatebox{75}{\textbf{\texttt{c880-CS640}}} & 
\rotatebox{75}{\textbf{\texttt{c880-RN640}}} & 
\rotatebox{75}{\textbf{\texttt{c880-CS1280}}} & 
\rotatebox{75}{\textbf{\texttt{c880-CS2560}}} & 
\rotatebox{75}{\textbf{\texttt{c880-RN2560}}} & 
\rotatebox{75}{\textbf{\texttt{c880-RN1280}}} & 
\rotatebox{75}{\textbf{\texttt{c880-RN320}}} & 
\rotatebox{75}{\textbf{\texttt{c880-SL1280}}} & 
\rotatebox{75}{\textbf{\texttt{c880-CS320}}} & 
\rotatebox{75}{\textbf{\texttt{c880-SL640}}} & 
\rotatebox{75}{\textbf{\texttt{c880-SL2560}}} & 
\rotatebox{75}{\textbf{\texttt{IBEX       }}} & 
\rotatebox{75}{\textbf{\texttt{MOR1KX     }}} & 
\rotatebox{75}{\textbf{\texttt{GPS        }}} 
\\
\midrule
261 & 209 & 251 & 357 & 345 & 252 & 345 & 212 & 2370 & 211 & 398 & 252 & 296 & 297 & 256 & 390 & 404 & 249 & 301 & 430 & 480 & 483 & 575 & 779 & 765 & 579 & 435 & 566 & 430 & 473 & 765 & 17K & 158K & 193K \\
\bottomrule
\end{tabular}
}
\end{table*}

\setlength{\tabcolsep}{\oldtabcolsep}

\subsection{Sizes of Target Netlists}\label{sec:appendix_design_sizes}
Table~\ref{tab:design_sizes} shows the sizes of the netlists we target. They range from a few hundred gates to a couple hundred thousand gates, showcasing that \myname{} works well across this wide spectrum of netlists.

\subsection{LLMs against Jplag and SIM}\label{sec:appendix_llms_against_Jplag_SIM}
The results from the evaluation of~\myname{} against the JPlag and SIM detection tools are shown below in Figures~\ref{fig:JPlag_boxplot} and~\ref{fig:Sim_boxplot3}, respectively. These figures demonstrate the hardware piracy performance (in terms of similarity scores) of each LLM for each of the 32 netlists.
Note that the distribution of similarity scores plotted for each netlist for each LLM are for the best mapping strategies for that netlist and LLM.
It is evident from both figures that, overall, we are able to successfully pirate the netlists and evade detection. For more detailed analysis and key takeaways, please refer to Secs.~\ref{sec:results_against_Jplag} and~\ref{sec:results_against_SIM}.

\begin{figure}[H]
    \centering
    \includegraphics[width=0.48\textwidth,trim={0.2cm 0.2cm 0.2cm 0.0cm},clip]{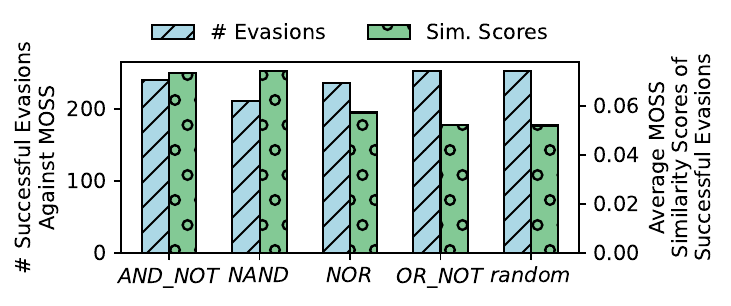}
    \caption{Performance of mapping strategies against MOSS.}
    \label{fig:mapping_strat_analysis_MOSS}
\end{figure}

\begin{figure}[H]
    \centering
    \includegraphics[width=0.48\textwidth,trim={0.2cm 0.2cm 0.2cm 0.0cm},clip]{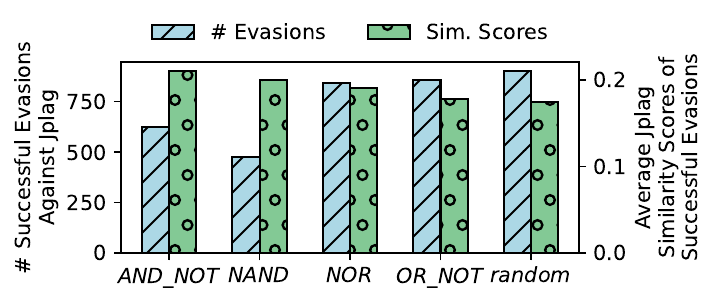}
    \caption{Performance of mapping strategies against Jplag.}
    \label{fig:mapping_strat_analysis_Jplag}
\end{figure}

\begin{figure}[H]
    \centering
    \includegraphics[width=0.48\textwidth,trim={0.2cm 0.2cm 0.2cm 0.0cm},clip]{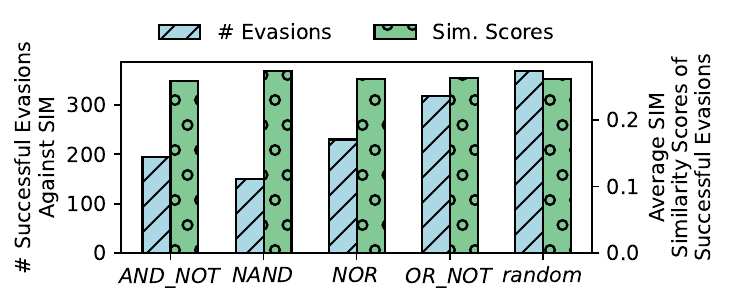}
    \caption{Performance of mapping strategies against SIM.}
    \label{fig:mapping_strat_analysis_SIM}
\end{figure}

\subsection{Analysis of MOSS, JPlag, and SIM Mapping Strategies}\label{sec:appendix_analysis_of_mapping_strategies_MOSS_Jplag_SIM}
Additional comparative evaluations between each mapping strategy used by~\myname{} are shown in Figures~\ref{fig:mapping_strat_analysis_MOSS},~\ref{fig:mapping_strat_analysis_Jplag}, and~\ref{fig:mapping_strat_analysis_SIM} against the MOSS, JPlag, and SIM piracy detection tools, respectively. For each mapping strategy (\textit{AND\_NOT}, \textit{NAND}, \textit{NOR}, \textit{OR\_NOT}, and \textit{random}), these figures compare the number of total number of successful evasions (from all LLMs and netlists) and the average similarity scores of those evasions.

It is evident across all figures that (similar to the results against GNN4IP in Sec.~\ref{sec:analysis_of_mapping_strategies_GNN4IP}) the \textit{random} mapping strategy results in the highest number of successful evasions. This finding is supported by the \textit{random} strategy demonstrating the lowest average similarity scores across most other strategies. 
For additional analysis of the superior performance of the \textit{random} strategy, see Sec.~\ref{sec:analysis_of_mapping_strategies_GNN4IP}, in which the GNN4IP detection tool is utilized.

\begin{table}
\caption{Number of successfully pirated netlists (out of 32) and average similarity scores (in brackets) against different detection tools.}
\label{tab:ablation_study}
\resizebox{0.5\textwidth}{!}{%
\begin{tabular}{ccccc}
\toprule
\textbf{Detection Tool} & GNN4IP~\cite{GNN4IP} & MOSS~\cite{moss_website} & Jplag~\cite{Jplag_github} & SIM~\cite{SIM_website} \\
\cmidrule{1-5}
\textbf{\myname{}\textbackslash Solution \ccfilledgreen{A}} & 0 (NA) & 0 (NA) & 0 (NA) & 0 (NA) \\
\textbf{\myname{}\textbackslash Solution \ccfilledgreen{B}} & 0 (NA) & 0 (NA) & 0 (NA) & 0 (NA) \\
\textbf{\myname{}\textbackslash Solution \ccfilledgreen{C}} & 32 (-0.75) & 32 (0.01) & 32 (0.20) & 7 (0.32) \\
\textbf{\myname{}} & 32 (-0.88) & 32 (0.01) & 32 (0.13) & 26 (0.27) \\
\bottomrule
\end{tabular}
}
\end{table}

\begin{figure*}
    \centering
    \includegraphics[width=\textwidth]{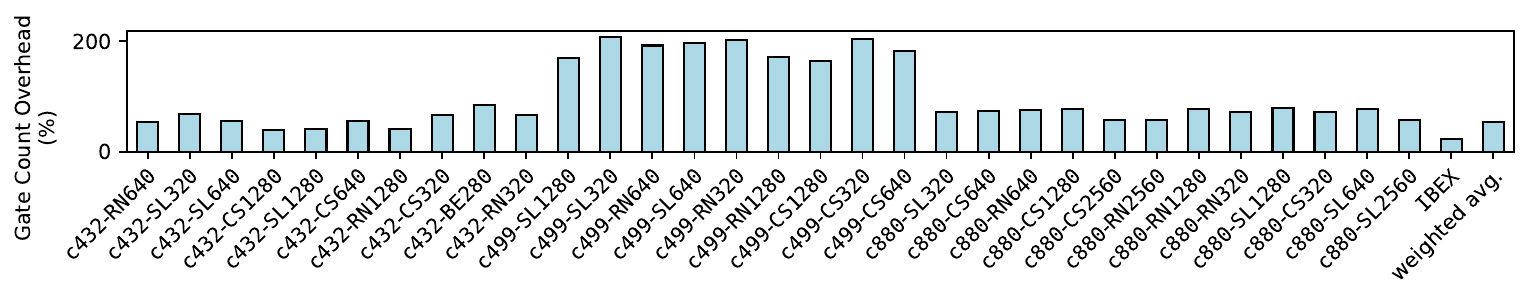}
    \caption{Gate count overheads of successfully pirated netlists against GNN4IP~\cite{GNN4IP}.}
    \label{fig:gate_cnt_overhead}
\end{figure*}

\begin{figure*}
    \centering
    \includegraphics[width=\textwidth]{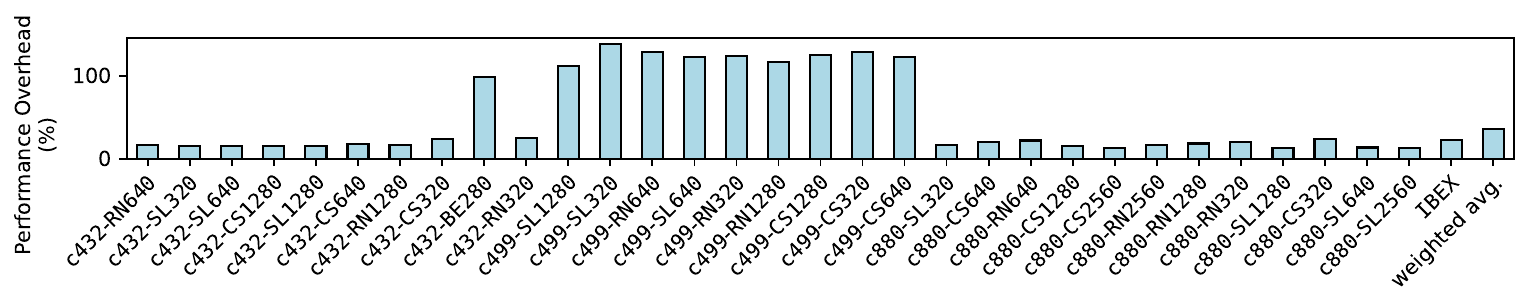}
    \caption{Performance overheads in terms of critical paths of successfully pirated netlists against GNN4IP~\cite{GNN4IP}.}
    \label{fig:performance_overhead}
\end{figure*}

\begin{figure*}[t]
    \centering
    \includegraphics[width=\textwidth,trim={0.2cm 0.2cm 0.2cm 0.2cm},clip]{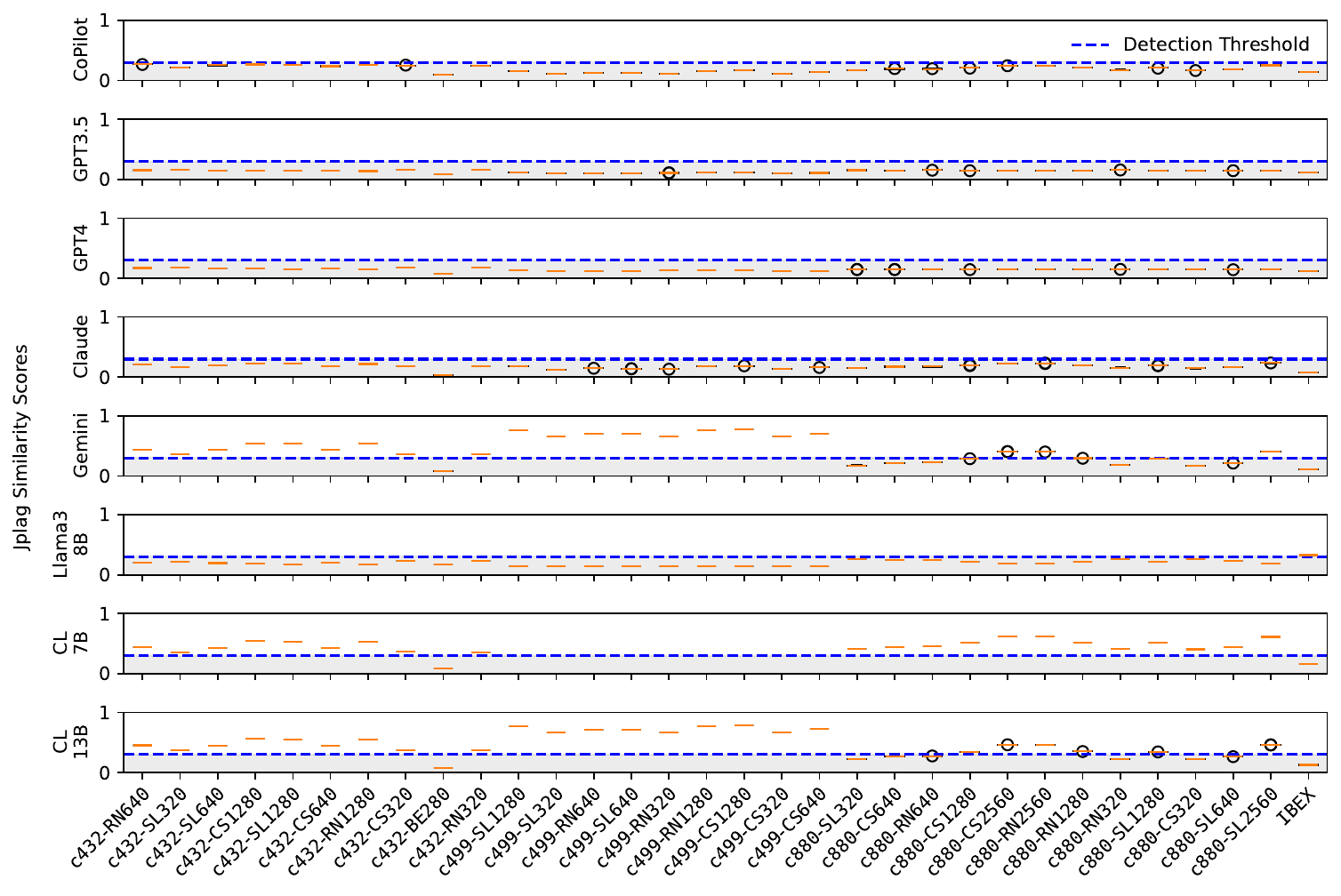}
    \caption{Distribution of Jplag similarity scores for different LLMs in \myname{}'s framework.}
    \label{fig:JPlag_boxplot}
\end{figure*}

\begin{figure*}[t]
    \centering
    \includegraphics[width=\textwidth,trim={0.2cm 0.2cm 0.2cm 0.2cm},clip]{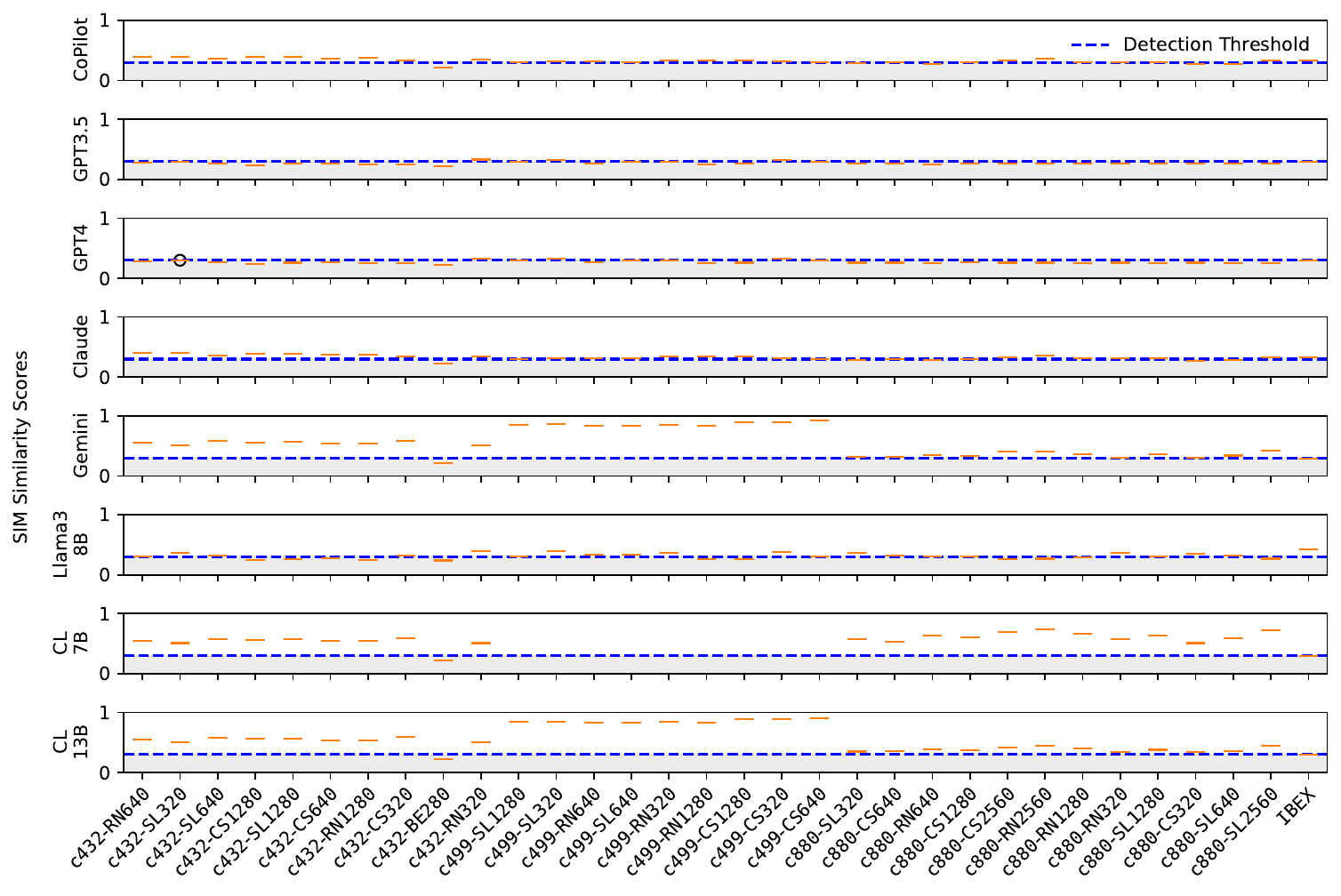}
    \caption{Distribution of SIM similarity scores for different LLMs in \myname{}'s framework.}
    \label{fig:Sim_boxplot3}
\end{figure*}

\subsection{Equivalence of Pirated Netlists}\label{sec:appendix_equivalence_of_pirated_netlists}
In addition to successfully pirating netlists that evade detection, \myname{} also guarantees functional equivalence of the pirated netlists. 
In order to guarantee equivalence, we incorporate exhaustive testing in the LLM-generated transformations. More specifically, we compare the original gate's output and the LLM-generated transformation's (i.e., gate/gates that would replace the original gate) output for all possible $2^n$ input vectors (assuming $n$ inputs), and consider a transformation successful only if the two outputs match for all input vectors. Since the pirated netlist is created from the original netlist by replacing only the original gate with a guaranteed equivalent transformation, the pirated netlist is guaranteed to be functionally equivalent to the original netlist.

Furthermore, we double-check the equivalence of the pirated netlists through formal verification. For each of the $31$ netlists from the GNN4IP repository, we check the equivalence of a successfully pirated netlist against each of the four detection tools.\footnote{We only verify the equivalence using the formal verification tool for the netlists from the GNN4IP repository because other netlists are too large to formally verify. However, \myname{} still guarantees equivalence of those larger netlists because of the exhaustive simulation-based testing explained above.} Then, we use the \textit{Cadence Conformal Equivalence Checker}~\cite{cadence_lec} to formally verify the equivalence of the pirated netlists with their original counterparts. Our results show that all sampled pirated netlists are classified as functionally equivalent to the corresponding original netlists.

\subsection{Ablation Study}\label{sec:appendix_ablation_study}
Here, we perform an ablation study to 
understand the contribution of the component to our overall \myname{} technique. Specifically, we evaluate the number of successfully pirated netlists without 
different solutions (Solution \ccfilledgreen{A}, Solution \ccfilledgreen{B}, and Solution \ccfilledgreen{C}). Recall that Solution \ccfilledgreen{B} can only be applied if Solution \ccfilledgreen{A} has been applied, because the former relies on the latter. Hence, removing Solution \ccfilledgreen{A} has to be accompanied with removal of Solution \ccfilledgreen{B}. In other words, during ablation study, removal of Solution \ccfilledgreen{A} means removal of Solution \ccfilledgreen{A} as well as Solution \ccfilledgreen{B}. Table~\ref{tab:ablation_study} shows the results, which demonstrate that while Solution \ccfilledgreen{C} is helpful in improving the performance of \myname{} (note the improvement in evading SIM and the average similarity scores for all detection tools in \myname{}), Solutions \ccfilledgreen{A} and \ccfilledgreen{B} are absolutely essential to achieve any success in pirating the netlists. This is because without Solutions \ccfilledgreen{A} or \ccfilledgreen{B}, several netlists exceed the context windows of the LLMs, and those that do not are still too complicated for LLMs to understand. Thus, overall, the relative importance of the different solutions for successfully pirating netlists is: Solution \ccfilledgreen{A}~$>$~Solution \ccfilledgreen{B}~$>>>$~Solution \ccfilledgreen{C}.

\subsection{Performance Overheads}\label{sec:appendix_performance_overheads}
In this section, we analyze the overheads incurred due to the modifications made by \myname{} that lead to successful piracy. Note that, since our threat model assumes piracy of pre-synthesized netlists, we don't assume access to a synthesis library, and instead estimate the performance of a given netlist based on two metrics: (i) the gate count, and (ii) the critical path of the netlist, i.e., the longest (measured by the number of gates) path from an input to an output. 

Figure~\ref{fig:gate_cnt_overhead} shows the gate count overheads of the successfully pirated netlists against GNN4IP. 
Note that the gate count overheads for most netlists and the weighted (by the number of gates) average overhead are $\approx50\%$. Such overheads are reasonable and a small price to pay for an attacker that can quickly and easily pirate valuable IPs.
Moreover, although analyzing the overheads in terms of gate counts is helpful, analyzing the overheads in terms of the critical paths is more important since the critical paths determine the operating frequency of the netlist. Figure~\ref{fig:performance_overhead} shows this overhead in terms of critical paths of successfully pirated netlists against GNN4IP.

Note that these overheads for most netlists, including \texttt{IBEX} as well as the weighted average overhead, are even lower, i.e., $\approx20\%$, thereby highlighting the small price an attacker has to pay for pirating valuable IPs. 
Another interesting observation from Figures~\ref{fig:gate_cnt_overhead} and~\ref{fig:performance_overhead} is that the \texttt{c499-*} netlists have the highest percentage overheads. This trend aligns with the trend in Figure~\ref{fig:GNN4IP_boxplot} where \texttt{c499-*} netlists have the highest similarity scores across all LLMs, meaning these netlists are very difficult to pirate. This observation prompted us to investigate these netlists in more detail, leading to the discovery that, unlike other netlists, the \texttt{c499-*} netlists contain an unusually large number of \texttt{XOR}/\texttt{XNOR} gates, and rewriting these gates using \texttt{NAND}/\texttt{NOR} operators requires at least $5$ gates, leading to large overheads. Thus, due to the large number of \texttt{XOR}/\texttt{XNOR} gates in them, successfully pirating \texttt{c499-*} netlists results in larger overheads.

\subsection{Details About Conversions form Boolean Formulas to Gate-level Netlists}\label{sec:appendix_conversion_from_boolean_formulas}

\begin{figure}[H]
\centering
\begin{lstlisting}[language=Verilog, label={listing:example_orig_gate}, caption={Example gate in original Verilog netlist.}, style=prettyverilog, belowskip=-0pt, aboveskip=20pt, firstnumber=1, linewidth=\linewidth]
nand U1 (n3, n1, n2rightbracket;
\end{lstlisting}
\end{figure}

\begin{figure}[H]
\centering
\begin{lstlisting}[language={}, label={listing:example_transformation}, caption={Example transformation from an LLM for the gate type in Listing~\ref{listing:example_orig_gate}.}, style=normaltext, belowskip=-0pt, aboveskip=20pt, firstnumber=1, linewidth=\linewidth]
N1 = AND(A1, A2)
Y = NOT(N1)
\end{lstlisting}
\end{figure}

\begin{figure}[H]
\centering
\begin{lstlisting}[language=Verilog, label={listing:example_pirated_gates}, caption={Gates in our pirated netlist corresponding to the original gate shown in Listing~\ref{listing:example_orig_gate}.}, style=prettyverilog, belowskip=-0pt, aboveskip=20pt, firstnumber=1, linewidth=\linewidth]
and U1 (tmp1, n1, n2rightbracket;
not U2 (n3, tmp1rightbracket;
\end{lstlisting}
\end{figure}

Once the transformations are generated using the LLMs (see Figure~\ref{fig:final_flow}), we employ a custom Python script to actually pirate a given original netlist.
Our script iterates over the gates in the given original netlist and replaces the gates with a netlist format of the LLMs' transformation in Boolean formulas format.
Listing~\ref{listing:example_orig_gate} shows an example gate in the original netlist, Listing~\ref{listing:example_transformation} shows an example transformation obtained from an LLM in a Boolean formula format (with template input names \texttt{A1} and \texttt{A2}, template output name \texttt{Y}, and intermediate variable \texttt{N1}), and Listing~\ref{listing:example_pirated_gates} shows the corresponding gate for our pirated netlist.

Also, as explained in Sec.~\ref{sec:appendix_equivalence_of_pirated_netlists}, to ensure functional correctness of the pirated netlist, we exhaustively test (by simulation of all possible $2^n$ input values for a gate with $n$ inputs) the LLM-generated Boolean formula for each unique original gate type, and compare the outputs. If all outputs match, the generated Boolean formula, and hence the gates in the pirated netlist, are guaranteed to be equivalent to the original gate(s).

\subsection{Discussion on Netlists Protected by Obfuscation}\label{sec:appendix_discussion_on_designs_protected_by_obfuscation}
Researchers have proposed hardware obfuscation techniques that involve ``locking'' circuits using key-controlled gates~\cite{roy2008epic,sengupta2020truly}. However, many of these techniques have been successfully compromised by previous research, raising significant doubts about the security of hardware obfuscation as a whole (see~\cite{engels2019end}). Given the vulnerabilities demonstrated in numerous hardware obfuscation techniques and the overarching concerns about their effectiveness, we do not consider them in our work since pirating such obfuscated netlists would only require an additional trivial step of un-obfuscating the netlist before using \myname{} to pirate it. To validate this claim, we obfuscated all $31$ netlists from the GNN4IP repository~\cite{hw2vec_github} using the popular AntiSAT~\cite{xie2018anti} obfuscation technique implemented in~\cite{anti_sat_implementation}. We then tested the security of the obfuscated netlists using a custom \textit{Python} implementation of the Signal Probability Skew (SPS) attack~\cite{yasin2017security}. Table~\ref{tab:deobfuscation_results} shows the results of the attack. It is evident that the attack successfully removes all obfuscated parts and recovers all original netlists.
\begin{table*}[t]
\caption{Success rate in de-obfuscating AntiSAT~\cite{xie2018anti}-protected netlists using the Signal Probability Skew (SPS)~\cite{yasin2017security} attack. 
}
\label{tab:deobfuscation_results}
\resizebox{\textwidth}{!}{%
\begin{tabular}{cccccccccccccccccccccccccccccccc}
\toprule & \rotatebox{75}{\textbf{\texttt{c432-RN640}}} & 
\rotatebox{75}{\textbf{\texttt{c432-SL320}}} & 
\rotatebox{75}{\textbf{\texttt{c432-SL640}}} & 
\rotatebox{75}{\textbf{\texttt{c432-CS1280}}} & 
\rotatebox{75}{\textbf{\texttt{c432-SL1280}}} & 
\rotatebox{75}{\textbf{\texttt{c432-CS640}}} & 
\rotatebox{75}{\textbf{\texttt{c432-RN1280}}} & 
\rotatebox{75}{\textbf{\texttt{c432-CS320}}} & 
\rotatebox{75}{\textbf{\texttt{c432-BE280}}} & 
\rotatebox{75}{\textbf{\texttt{c432-RN320}}} & 
\rotatebox{75}{\textbf{\texttt{c432-SL1280}}} & 
\rotatebox{75}{\textbf{\texttt{c499-SL320}}} & 
\rotatebox{75}{\textbf{\texttt{c499-RN640}}} & 
\rotatebox{75}{\textbf{\texttt{c499-SL640}}} & 
\rotatebox{75}{\textbf{\texttt{c499-RN320}}} & 
\rotatebox{75}{\textbf{\texttt{c499-RN1280}}} & 
\rotatebox{75}{\textbf{\texttt{c499-CS1280}}} & 
\rotatebox{75}{\textbf{\texttt{c499-CS320}}} & 
\rotatebox{75}{\textbf{\texttt{c499-CS640}}} & 
\rotatebox{75}{\textbf{\texttt{c880-SL320}}} & 
\rotatebox{75}{\textbf{\texttt{c880-CS640}}} & 
\rotatebox{75}{\textbf{\texttt{c880-RN640}}} & 
\rotatebox{75}{\textbf{\texttt{c880-CS1280}}} & 
\rotatebox{75}{\textbf{\texttt{c880-CS2560}}} & 
\rotatebox{75}{\textbf{\texttt{c880-RN2560}}} & 
\rotatebox{75}{\textbf{\texttt{c880-RN1280}}} & 
\rotatebox{75}{\textbf{\texttt{c880-RN320}}} & 
\rotatebox{75}{\textbf{\texttt{c880-SL1280}}} & 
\rotatebox{75}{\textbf{\texttt{c880-CS320}}} & 
\rotatebox{75}{\textbf{\texttt{c880-SL640}}} & 
\rotatebox{75}{\textbf{\texttt{c880-SL2560}}} 
\\
\midrule
Key Size & 64 & 64 & 64 & 64 & 64 & 64 & 64 & 64 & 64 & 64 & 64 & 64 & 64 & 64 & 64 & 64 & 64 & 64 & 64 & 64 & 64 & 64 & 64 & 64 & 64 & 64 & 64 & 64 & 64 & 64 & 64  \\
\midrule
Success Rate (\%) & 100 & 100 & 100 & 100 & 100 & 100 & 100 & 100 & 100 & 100 & 100 & 100 & 100 & 100 & 100 & 100 & 100 & 100 & 100 & 100 & 100 & 100 & 100 & 100 & 100 & 100 & 100 & 100 & 100 & 100 & 100 \\
\midrule
Runtime (s) & $<0.1$ & $<0.1$ & $<0.1$ & $<0.1$ & $<0.1$ & $<0.1$ & $<0.1$ & $<0.1$ & $<0.1$ & $<0.1$ & $<0.1$ & $<0.1$ & $<0.1$ & $<0.1$ & $<0.1$ & $<0.1$ & $<0.1$ & $<0.1$ & $<0.1$ & $<0.1$ & $<0.1$ & $<0.1$ & $<0.1$ & $<0.1$ & $<0.1$ & $<0.1$ & $<0.1$ & $<0.1$ & $<0.1$ & $<0.1$ & $<0.1$ \\
\bottomrule
\end{tabular}
}
\end{table*}

\subsection{Example Transformations}
\label{sec:appendix_example_transformations}
\begin{figure*}[t]
    \centering
    \includegraphics[width=\textwidth]{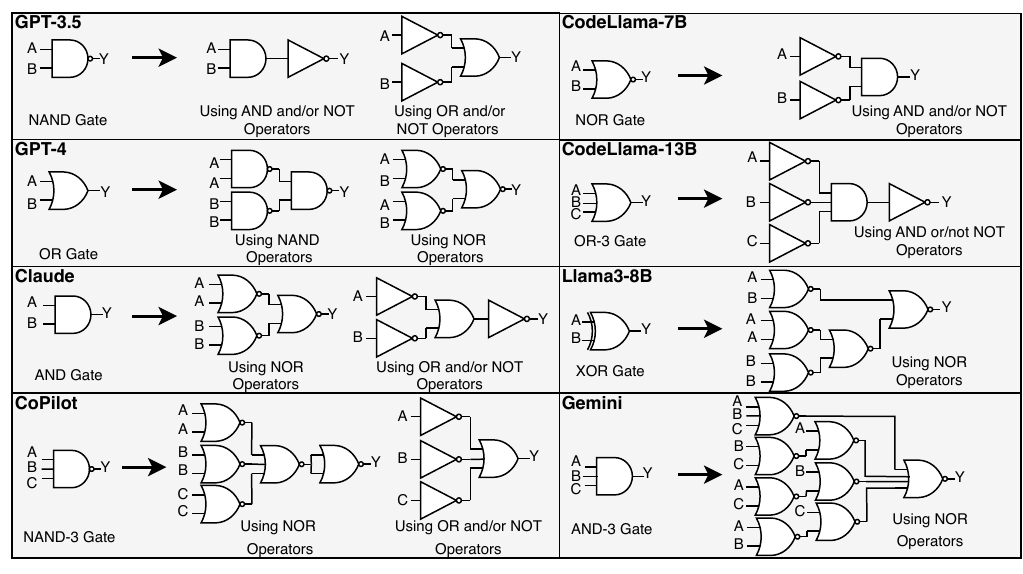}
    \caption{Example netlist transformations from LLMs.}
    \label{fig:example_circuit_transformations}
\end{figure*}

Figure~\ref{fig:example_circuit_transformations} illustrates some example valid transformations generated by the eight LLMs. The illustrations show that LLMs are able to generate a variety of transformations, including simple ones, such as  \texttt{NAND} gate using [\texttt{AND}, \texttt{NOT}] Boolean operators) and complicated ones, such as \texttt{XOR} gate using [\texttt{NOR}] Boolean operator.

\end{document}